\documentclass[twocolumn, notitlepage, floatfix, superscriptaddress, showpacs, showkeys, prx, longbibliography]{revtex4-1}
\usepackage{amsmath,amssymb,amsfonts,bm, makecell}
\usepackage{graphicx}
\usepackage{tabularx}
\usepackage{hyperref}
\usepackage{verbatim}

\usepackage{color,xcolor,soul}
\setstcolor{red}

\newcommand{\nubar}{\bar{\nu}}
\newcommand{\isample}{{(i)}}
\newcommand{\bart}{{\bar t}^\isample}
\usepackage{kotex}

\begin{document}

\title{Unified framework for hybrid percolation transitions based on microscopic dynamics}

\author{Hoyun Choi}
\affiliation{CTP and Department of Physics and Astronomy, Seoul National University, Seoul 08826, Korea}


\author{Y.S. Cho}
\affiliation{Department of Physics, Jeonbuk National University, Jeonju 54896, Korea}

\author{Raissa D'Souza}
\affiliation{Department of Computer Science and Department of Mechanical and Aerospace Engineering, University of California, Davis, California 95616, USA}
\affiliation{Santa Fe Institute, Santa Fe, New Maxico 87501, USA}

\author{János Kertész}
\affiliation{Department of Network and Data Science, Central European University, Quellen strasse 51, 1100 Vienna, Austria}

\author{B. Kahng}
\email{bkahng@kentech.ac.kr}
\affiliation{CCSS, KI for Grid Modernization, Korea Institute of Energy Technology, Naju, Jeonnam 58217, Korea}

\begin{abstract}
A hybrid percolation transition (HPT) exhibits both discontinuity of the order parameter and critical behavior at the transition point.
Such dynamic transitions can occur in two ways: by cluster pruning with suppression of loop formation of cut links or by cluster merging with suppression of the creation of large clusters.
While the microscopic mechanism of the former is understood in detail, a similar framework is missing for the latter.
By studying two distinct cluster merging models, we uncover the universal mechanism of the features of HPT-s at a microscopic level.
We find that these features occur in three steps: (i) medium-sized clusters accumulate due to the suppression rule hindering the growth of large clusters, (ii) those medium size clusters eventually merge and a giant cluster increases rapidly, and (iii) the suppression effect becomes obsolete and the kinetics is governed by the Erd\H{o}s-R\'enyi type of dynamics.
We show that during the second and third period, the growth of the largest component must proceed in the form of a Devil's staircase.
We characterize the critical behavior by two sets of exponents associated with the order parameter and cluster size distribution, which are related to each other by a scaling relation.
Extensive numerical simulations are carried out to support the theory where a specific method is applied for finite-size scaling analysis to enable handling the large fluctuations of the transition point.
Our results provide a unified theoretical framework for the HPT.
\end{abstract}

\maketitle

\section{Introduction}
Hybrid phase transitions, sometimes referred to as mixed-order transitions, exhibit features of both first- and second-order transitions at the same transition point~\cite{hpt_intro1, hpt_intro2}.
Recently, hybrid phase transitions have been observed in various models and systems, including models of wetting~\cite{blossey1995diverging} and DNA denaturation~\cite{kafri2000dna}, spin models with long-range interactions in one dimension~\cite{bar2014mixed, hpt_eq1, hpt_eq2, hpt_eq3}, and short-range competing interactions, such as the Ashkin-Teller model, in scale-free networks at critical endpoints~\cite{jang2015ashkin}, glass and jamming models~\cite{toninelli2006jamming}, epidemic contagion models on complex networks~\cite{janssen2016first,choi2017mixed, hpt_epid1}, synchronization~\cite{pazo2005thermodynamic, epsync, epsync2, hpt_sync1, hpt_sync2, hpt_sync3}, and percolation~\cite{panagiotou2011explosive,buldyrev2010catastrophic, baxter, kcore, kchoi2021, hpt_perc1, hpt_perc2, hpt_perc3, hpt_perc4} models.
In addition, a hybrid phase transition was experimentally demonstrated in a colloidal crystal~\cite{alert2017mixed}. However, the understanding of the microscopic mechanism underlying this type of phase transition is far from being complete~\cite{nuno, hpt_rev1, hpt_rev2}.

In percolation on a network, a geometric phase transition, the control parameter $t=L/N$ is the density of occupied (active) links with $L$ being the number of occupied links, and $N$ the number of nodes. The order parameter $m$ is the probability that a node belongs to a giant cluster as $N\to\infty$~\cite{er, ep, perc_intro}. In Bernoulli percolation~\cite{stauffer}, where the occupation probability of the links is independent, there is usually (e.g., on regular lattices in dimensions higher than one or on a complete graph) a second-order or continuous transition from the non-percolating to the percolating phase characterized by the emergence of a giant component.
The transition is accompanied by critical phenomena, like the power law behavior of thermodynamic quantities near the critical point, which are characterized by critical exponents, e.g., $\beta, \gamma, \nu$, for the order parameter, the susceptibility, and the correlation length, respectively~\cite{stauffer, perc_criticality}. The percolation transition can also be described in terms of the probability distribution; $p_s=s^{1-\tau}\hat{f}(s/s^*)$ with $s^*\sim |t-t_c|^{-1/\sigma}$, where $p_s\equiv sn_s$ is the probability that a node belongs to a cluster of size $s$, $n_s$ is the normalized number of clusters of size $s$, and $\hat{f}$ is a scaling function. Of all the exponents, only two are independent.

Here, we focus on hybrid percolation transitions (HPTs) on complex networks.
In an HPT, the order parameter is discontinuous at $t_c^-$ when the transition point $t_c$ is approached from below; and it exhibits critical behavior on the other side as $t \to t_c^+$ from above. This behavior is expressed as follows:
\begin{equation}
m(t)=\left\{
\begin{array}{lr}
0                      & ~{\rm for}~~ t < t_c,    \\
m_c+r(t-t_c)^{\beta_m} & ~{\rm for}~~ t \geq t_c,
\end{array}
\right.
\label{eq:hybridPT_order}
\end{equation}
where $m_c = m(t_c)$; $r$ is a $t$-independent constant; and $\beta_m$ is a critical exponent associated with the order parameter.

Bernoulli percolation can be considered via two possible methods: Either the links are occupied with a probability corresponding to $t$ (static picture), or the links are occupied one by one until their density reaches $t$ (kinetic picture~\cite{souza2015anomalous,ben2005kinetic}) - and these two are equivalent in the thermodynamic limit.
Here we are dealing with dynamic models with correlated occupations, so we have to apply the kinetic representation, which has the additional advantage that we can trace the evolution of the characteristic quantities.

HPTs can occur both in cluster pruning ($t\to t_c^+$) and cluster merging ($t\to t_c^-$) processes. For cluster pruning ($t\to t_c^+$), it is established that the discontinuity of the order parameter is related to a dynamically evolving metastable state while the critical behavior is induced by cascading failures with the underlying universal mechanism identified as a critical branching process~\cite{lee2017universal,zhou2014}.
To characterize the critical behaviors, two sets of critical exponents were introduced~\cite{mcc}: one is associated with the order parameter (i.e., the giant cluster) and the other is associated with the power-law behavior of the avalanche size distribution; moreover, a scaling law relates these two sets of critical exponents.

For the case of the cluster-merging process such a universal mechanism has yet to be identified. A challenge is that the dynamic evolution of the system starts from $t=0$, thus the critical behavior of HPT is affected by the process leading to the jump of the order parameter. The following natural questions arise: Is there a universal microscopic mechanism behind the HPT in cluster merging processes? Can the critical behaviors of the order parameter and the clusters be characterized by critical exponents like for the cluster pruning processes? Is there a relationship between the critical behaviors of the cluster pruning and the cluster merging processes? These are the questions we address in this paper.

The paper is organized as follows.
In Sec.~\ref{sec:kcore}, we review the results on the microscopic mechanism of the critical phenomena in the HPT due to cluster-pruning processes.
In Sec.~\ref{subsec:rer} and~\ref{subsec:bfw} we reconsider the critical behavior of the $r$-ER model, and introduce and consider the $m$-BFW model from a common perspective; we discuss the universal scaling relation that unifies the critical behaviors in the cluster-merging and cluster-pruning processes.
In Sec.~\ref{sec:conclusion}, we summarize the results and discuss their implications.

\section{HPT in cluster-pruning process}    \label{sec:kcore}
In this section we present the established results for the HPT in the cluster pruning process~\cite{mcc}, for instance, the critical behavior of $k$-core percolation with $k=3$ on the Erd\H os-R\'enyi (ER) graph~\cite{er}. In this case, the kinetics starts from a supercritical state at $t \gg t_c$. The $k$-core~\cite{Chalupa1979bootstrap,kcore_mendes,jo_mj} is a subgraph of the network in which each node has at least $k$ degrees. To obtain an initial $k$-core subgraph, an ER network is generated at $t \gg t_c$, and then all nodes with a degree of less than $k$ are iteratively removed along with their links. From the resulting stable $k$-core subgraph, a randomly selected node is removed along with its links. This removal may lead to other nodes getting knocked out of the $k$-core in an avalanche of activity, reducing the size of the $k$-core. The amount by which the size is reduced corresponds to avalanche size.
As this process is repeated, the order parameter decreases gradually and exhibits critical behavior following Eq.~\eqref{eq:hybridPT_order}.
At $t_c^+$, the order parameter suddenly drops from a finite value to zero.

The features of this critical behavior are as follows.
(i) The critical exponent can be derived analytically as $\beta_m=1/2$~\cite{mcc}, which is universal regardless of the different models, such as the cascade failure model of interdependent networks~\cite{buldyrev2010catastrophic}.
(ii) The critical behavior for $t > t_c$ is characterized by two sets of critical exponents: one set $\{\beta_m$, $\gamma_m$, $\bar \nu_m\}$ is associated with the order parameter, and the other set $\{\tau_a, \sigma_a, \beta_a, \gamma_a, \nubar_a\}$ is associated with the avalanche size distribution.
For instance, the fluctuations of the order parameter over different configurations are expressed by the susceptibility, which is defined as $\chi_m \equiv N(\langle m^2 \rangle - \langle m \rangle^2) \sim (t-t_c)^{-\gamma_m}$.
However, the mean avalanche size is expressed as $\chi_a \equiv \sum_s^\prime s p_s(t) /\sum_s^\prime p_s(t)\sim (t-t_c)^{-\gamma_a}$, where the prime denotes the summation over finite avalanche sizes, and $p_s(t)$ denotes the avalanche size distribution at $t$.
As $\gamma_m\approx 1.0$ and $\gamma_a = 1/2$, the two $\gamma$-s are different (though respectively universal for the different cluster pruning models on the ER model~\cite{mcc}).
The exponents $\tau_a$, $\sigma_a$, and $\bar \nu_a$ characterize the avalanche size distribution: $n_s\sim s_a^{-\tau_a}f(s_a/s_a^*)$, where $s_a$ denotes the avalanche size, $f$ is a scaling function, and $s_a^*$ is the characteristic avalanche size, which behaves as $\sim (t-t_c)^{-1/\sigma_a}$ for $N\to \infty$.
Its finite size behavior at the transition point is given by $s_a^*\sim N^{1/\sigma_a\bar \nu_a}$.
(iii) The critical exponents of the individual set satisfy the scaling relations; however, the two sets are not independent, but they are related through a conservation relation~\cite{mcc}:
\begin{equation}
    m(t)+\int_t^{t_0} \sum_s^\prime s p_s(t_1)d t_1=1, \label{eq:pruning}
\end{equation}
leading to
\begin{equation}
    \gamma_a=1-\beta_m.
    \label{eq:gamma_a_beta_m}
\end{equation}
This scaling relation is universal for the cascading failure models.
Later we will show that the relation $\eqref{eq:gamma_a_beta_m}$ is also valid for the HPT in the cluster-merging process.

\section{HPT in cluster-merging process}    \label{sec:merging}
\subsection {Modified $r$-ER model}   \label{subsec:rer}
\begin{table*}[!tb]
\renewcommand{\arraystretch}{1.2}
\centering
\setlength{\tabcolsep}{1.5em}
\begin{tabular}{*{7}{c}}
        \hline\hline
        $g$ & $\tau_s$        & $\sigma_s$      & $\beta_s$       & $\nubar_s$        & $\gamma_s$      & $\gamma_s^\prime$ \\
        \hline
        0.2 & $2.08 \pm 0.04$ & $0.99 \pm 0.05$ & $0.09 \pm 0.05$ & $1.10 \pm 0.10$   & $0.91 \pm 0.05$ & $1.03 \pm 0.02$   \\
        0.5 & $2.18 \pm 0.04$ & $0.96 \pm 0.05$ & $0.19 \pm 0.05$ & $1.24 \pm 0.11$   & $0.83 \pm 0.05$ & $1.10 \pm 0.03$   \\
        0.8 & $2.25 \pm 0.04$ & $0.89 \pm 0.05$ & $0.29 \pm 0.07$ & $1.42 \pm 0.13$   & $0.81 \pm 0.05$ & $1.15 \pm 0.05$   \\
        \hline\hline
        $g$ & $\beta_m$       & $\nubar_m$      & $\gamma_m$      & $\gamma_m^\prime$ & $\zeta$         & $\zeta^\prime$    \\
        \hline
        0.2 & $0.09 \pm 0.05$ & $1.05 \pm 0.10$ & $0.90 \pm 0.10$ & $2.06 \pm 0.05$   & $1.04 \pm 0.05$ & $1.89 \pm 0.10$   \\
        0.5 & $0.21 \pm 0.05$ & $1.25 \pm 0.15$ & $0.83 \pm 0.10$ & $2.30 \pm 0.10$   & $1.02 \pm 0.05$ & $1.76 \pm 0.13$   \\
        0.8 & $0.32 \pm 0.05$ & $1.40 \pm 0.15$ & $0.76 \pm 0.10$ & $2.55 \pm 0.10$   & $1.00 \pm 0.07$ & $1.65 \pm 0.13$   \\
        \hline\hline
    \end{tabular}
    \caption{
        List of numerical values of the exponents for HPT in the modified $r$-ER model.
        The numerical values of all critical exponents except $\zeta$ and $\zeta^\prime$ are newly obtained.
        They are consistent within the error bars with those presented in Ref.~\cite{cho2016hybrid}.
        The exponents values of $\zeta$ and $\zeta^\prime$ are adopted from Ref.~\cite{park2019interevent}.
        The definitions of each critical exponent and the scaling relations among them are listed in Appendix~\ref{app:exponent}.
    }   \label{table1}
\end{table*}

We next present results for the HPT due to cluster merging for two distinct models and establish the underlying universal mechanism.
When HPT occurs as a consequence of a cluster merging process the control parameter moves in the opposite direction as compared to cluster pruning processes from small to large values of $t$: the order parameter first jumps, and then the critical behavior is observed. Thus, the critical behavior is affected by the dynamics that occur leading to the jump of the order parameter as $t\to t_c^-$. The first model we study is the restricted ER model (denoted as the $r$-ER model).
This model was originally proposed to study a discontinuous percolation transition~\cite{panagiotou2011explosive}; here we consider a slightly modified version, the modified $r$-ER model, which was shown to exhibit an HPT~\cite{cho2016hybrid} (see Appendix~\ref{app:diff_rER}).

In this model, the cluster coalescence dynamics begin with $N$ isolated nodes.
At each time step, the clusters are ranked by size and partitioned into one set of small and one set of large clusters, denoted as $A$ and $B$, respectively.
In the original $r$-ER model, the $g$ fraction of nodes contained in the smallest clusters is assigned to set $A$, where $g$ ($0 < g < 1$) is a model parameter.
The remaining fraction of nodes is assigned to set $B$.
Hence one cluster can have nodes belonging to both sets (see Fig.~\ref*{fig:a1}).
We make the modification that when a cluster has nodes that span both sets $A$ and $B$, all nodes in the cluster are regarded as elements of set $A$.
From there we proceed as with the original $r$-ER model.
Two nodes are randomly selected from different clusters for cluster coalescence: one is from the entire system and the other is only from set $A$.
Thus, two nodes in set $B$ cannot be linked and this restriction rule suppresses the growth of large clusters. Accordingly, large clusters are rarely generated, whereas medium-sized clusters are generated in abundance.
As $t$ approaches $t_c$, medium-sized clusters are more likely to be merged, and the large cluster size rapidly increases.
In the limit $N\to \infty$, the growth rate of the giant cluster with respect to $t$ becomes infinity, and the change of the order parameter becomes discontinuous.

In finite systems, the cluster-merging process is characterized by three time regimes: $[0,t_a]$, $[t_a,t_g]$, and $[t_g, t_c]$~\cite{park2019interevent} (see Appendix B). In the early regime $t\in [0,t_a]$, due to the suppression effect, a bump is formed in the cluster-size distribution and its size increases first with $t$ corresponding to a large number of medium-size clusters. Here $t_a$ is defined as the time at which the bump size reaches its maximum. The bump is consistent with a ``powder keg" as first discussed by Friedman and Landsberg~\cite{Friedman2009Construction}. In the intermediate regime $[t_a, t_g]$, the bump size shrinks as medium-sized clusters merge with the giant cluster, and the size of the giant cluster increases rapidly.

At $t_g$, partition $B$ is fully occupied by the giant cluster alone; that is, $m(t_g)=1-g$. Hence, for $t \ge t_g$, the giant component must contain nodes in set $A$, thus the partition into two categories loses its meaning. In the regime $[t_g, t_c]$, the cluster-merging kinetics proceeds according to the ER rule~\cite{cho_rate}; however, the initial configuration of the ER kinetics is not the standard initial condition of isolated nodes, but the specific cluster configuration at $t_g$. The size distribution of finite clusters exhibits a power-law decay in a small cluster-size region with exponent $\tau_s$, which is not the value of the ER (or mean-field) class $\tau_s^{\rm{ER}}=5/2$ but depends on the model parameter $g$. When the system reaches $t_c$ the bump is completely eliminated, and finite clusters have a size distribution following a
Stauffer-type cluster scaling form with exponent $\tau_s$, i.e., $n_s\sim s^{-\tau_s}\exp(-s/s^*)$, where $s^*$ is a characteristic cluster size due to the finite-size effect:
\begin{equation}
    s^*\sim N^{1/\sigma_s\bar \nu_s},
    \label{eq:finite_size_scaling}
\end{equation}
with system size $N$.
The cluster scaling formula is valid also for $t_g < t < t_c$ with
\begin{equation}
    s^* \sim (t_c-t)^{-1/\sigma_s^\prime},
    \label{eq:char_cluster_scaling}
\end{equation}
as long as $s^*$ is smaller than the value given by (\ref*{eq:finite_size_scaling}).
The critical behavior of the clusters can be described using exponents $\{\tau_s, \sigma_s, \nubar_s \}$.
Since the merging of finite clusters continues to follow the ER mechanism also for $t>t_c$, the critical behavior in this regime is characterized by two sets of critical exponents: one set $\{\beta_m$, $\gamma_m$, $\bar \nu_m\}$ is associated with the order parameter, and the other set $\{\tau_s, \sigma_s, \beta_s, \gamma_s, \nubar_s\}$ is associated with the size distribution of finite clusters.

\begin{figure}[!htb]
    \centering
    \includegraphics[width=1.00\linewidth]{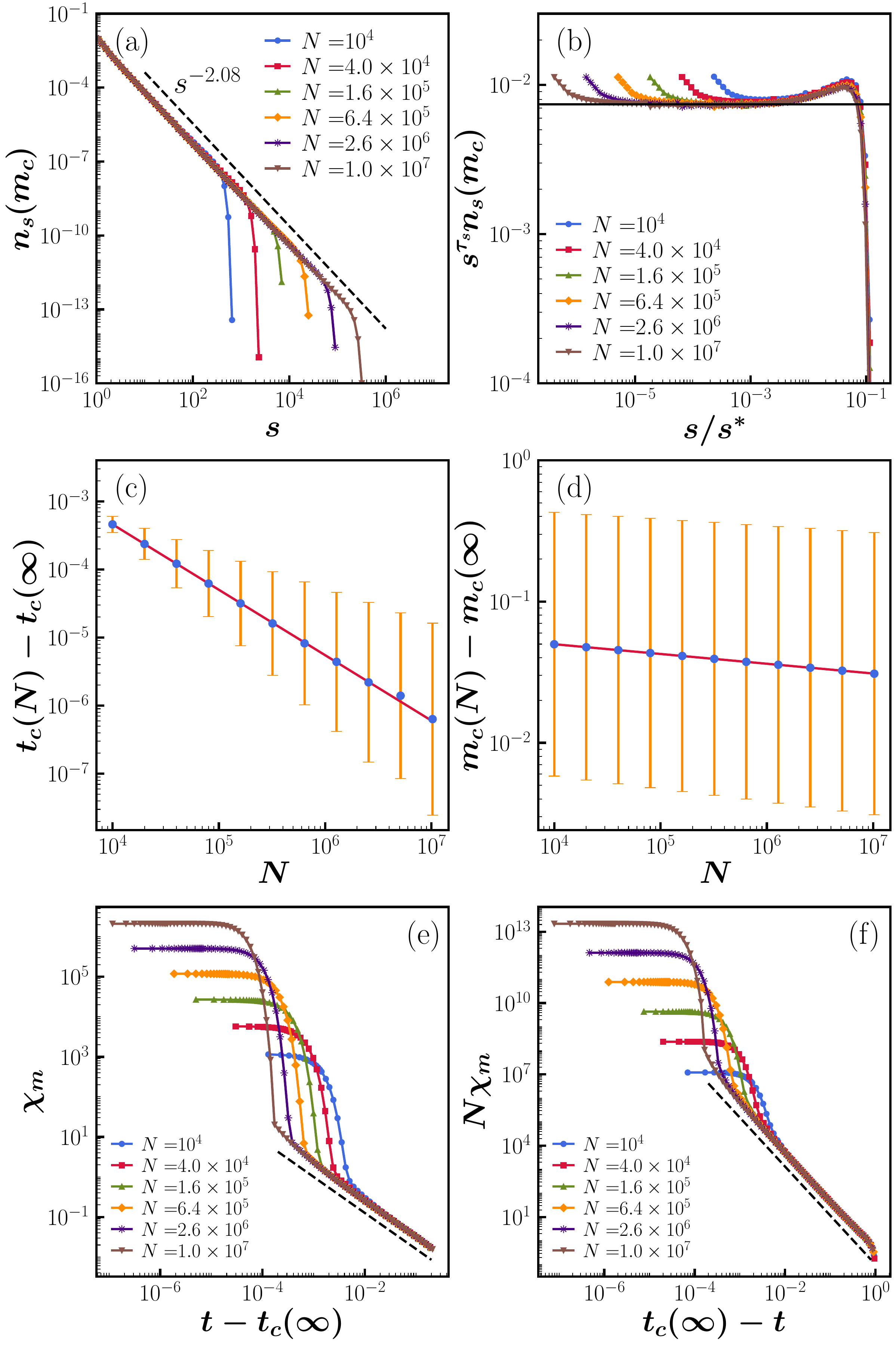}
    \caption{
    Measurement of the critical exponents of the modified $r$-ER model for $g=0.2$.
    (a) Plot of $n_s(m_c)$ vs $s$. $n_s(m_c)$ is measured just after $m$ exceeds $m_c$ for different system sizes $N/10^4 = 2^0, 2^2, 2^4, 2^6, 2^8,~{\rm and}~2^{10}$ from the left.
    Here, $m_c$ of each $N$ is taken as a value immediately after the bump of $n_s(m_c)$ disappears.
    A dashed line with slope $-2.08$ is drawn to guide the eye.
    (b) Scaling plot of $s^{\tau_s}n_s(m_c)$ vs $s/s^*$ for different sizes $N$ in (a), where $\tau_s = 2.08$ and $s^*=N^{1/\sigma_s\nubar_s}$ with $\sigma_s=0.99$ and $\nubar_s = 1.10$.
    (c) Plot of $t_c(N)-t_c(\infty)$ vs $N$ with error bars for $t_c(N) > t_c(\infty)$, where $t_c(N)$ is the configuration average of $t$ just after $m(t)$ exceeds $m_c(N)$.
    The slope of the solid line is $-1/1.05$, which supports that $\nubar_m \approx 1.05$ by the relation $t_c(N)-t_c(\infty) \sim N^{-1/\nubar_m}$.
    (d) Plot of $m_c(N)-m_c(\infty)$ vs $N$ with error bars.
    The slope of the solid line is $-0.07$, which supports that $\beta_m/\overline{\nu}_m \approx 0.07$ by the relation $m_c(N)-m_c(\infty) \sim N^{-\beta_m/\nubar_m}$.
    (e) Plot of $\chi_m(t)$ vs $(t-t_c(\infty))$ for $t > t_c(\infty)$ using $N/10^4 = 2^0, 2^2, 2^4, 2^6, 2^8,~{\rm and}~2^{10}$ from the right, where $\chi_m(t)=N(\langle m(t)^2 \rangle-\langle m(t)\rangle^2)$. The slope of the dashed line is $-0.90$, which supports that $\gamma_m \approx 0.90$ by the relation $\chi_m(t) \sim (t-t_c(\infty))^{-\gamma_m}$.
    (f) Plot of $N\chi_m(t)$ vs $(t_c(\infty)-t)$ for $t < t_c(\infty)$ using $N/10^4 = 2^0, 2^2, 2^4, 2^6, 2^8,~{\rm and}~2^{10}$ from the right. The slope of the dashed line is $-2.06$, which supports that $\gamma_m^\prime \approx 2.06$ by the relation $N\chi_m(t) \sim (t_c(\infty)-t)^{-\gamma_m^\prime}$.
    }  \label{fig:RER_exponents}
\end{figure}

We note that even though the evolution proceeds under the ER rule in regime $[t_g, t_c]$, the critical exponent $\tau_s$ does not have the ER value.
Instead, it is determined by the early time process.
As clusters are merged and their rankings are updated, they may move back and forth between the two sets.
The inter-event time (denoted as $z_i$) is defined as the time interval between two set-changing events of the cluster containing node $i$~\cite{park2019interevent}.
Its distribution $P_I(z)$ comprising these inter-event times over all nodes during the time interval $[0,t_a]$ exhibits a power-law decay, $P_I(z)\sim z^{-\zeta}$.
It was derived~\cite{park2019interevent} that the exponent $\zeta$ is related to the exponents $\tau_s$ and $\sigma_s$ as $\zeta=4-(\tau_s+\sigma_s)$.
Because the ER dynamics are dominant in $[0,t_a]$ and the ER values are $\tau_s=5/2$ and $\sigma_s=1/2$, $\zeta=1$ is predicted.
Indeed, numerical simulations produce $\zeta=1$, which is insensitive to model parameter $g$.

The exponent of $P_I(z)$ during the time interval $[t_a, t_g]$ is denoted by $\zeta^\prime$.
Since in this interval, $s^*$ can be considered as constant, $\sigma_s^\prime=0$.
Thus, the above relationship changes to $\zeta^\prime=4-\tau_s$~\cite{park2019interevent}.
The dependence of $\zeta^\prime$ on $g$ indicates that the critical exponent $\tau_s$ also depends on it.

We present numerical values of the two sets of critical exponents for different values of $g$ in Table.~\ref{table1}.
The critical exponents \{$\tau_s, \sigma_s, \gamma_s, \gamma_s^\prime, \beta_m$\} for different $g$ were directly measured from numerical simulations.
Then, $\nubar_s$ and $\beta_s$ are calculated using the scaling relations $\beta_s = (\tau_s-2)/\sigma_s$ and $\nubar_s = (\tau_s-1)/\sigma_s$, respectively.
When the system reaches $t_c$ ($m$ reaches $m_c$), we check that $s^{\tau_s}n_s$ vs $s/s*=sN^{-1/\sigma_s\nubar_s}$ for different system sizes are well collapsed using $\sigma_s$ and $\nubar_s$ within the error range of numerical values, irrespective of $g$ (See Fig.~\ref{fig:RER_exponents}(a) and (b) for the result of $g=0.2$.).
$\nubar_m, \gamma_m$, and $\gamma_m^\prime$ are obtained from simulation data as shown in Fig.~\ref{fig:RER_exponents}(c)--(f).
The values of $\zeta$ and $\zeta'$ are taken from \cite{park2019interevent}.

Next, we check whether the numerical values of exponents in Table.~\ref{table1} satisfy the scaling relations.
For $g=0.2$, the exponent of the order parameter $\beta_m$ defined in the formula $m-m_c \sim (t-t_c)^{\beta_m}$ is directly measured as $\beta_m=0.09\pm 0.05$.
This value is consistent with $\beta_s=0.09\pm 0.05$ obtained using the relation $(\tau_s-2)/\sigma_s$.
Secondly, using $\tau_s$ and $\sigma_s$, $\gamma_s$ is obtained using the scaling relation $(3-\tau_s)/\sigma_s$ as $\gamma_s=0.93 \pm 0.09$.
This value is consistent with the value $\gamma_s=0.91\pm 0.05$ directed measured in~\cite{cho2016hybrid} using the formula of the average size of finite clusters $\langle s \rangle \sim (t-t_c)^{-\gamma_s}$, where $\langle s \rangle \equiv \sum'_s s^2n_s(t)/\sum'_s sn_s(t)$.
Therefore, the scaling relation (\ref{eq:gamma_a_beta_m}) between the two types of critical exponents is valid within the error range as $\gamma_s + \beta_m = 1.00 \pm 0.10$.
Note that we used $\gamma_s$ instead of $\gamma_a$.
Finally, the two exponents related to the inter-event time distribution $\zeta = 1.04 \pm 0.05$ and $\zeta' = 1.89 \pm 0.10$ satisfy the relations $\zeta = 1$ and $\zeta' = 4-\tau_s = 1.92 \pm 0.04$ within the error ranges.
We also check the scaling relations for $g=0.5$, finding that they are valid within the error ranges.
However, for $g=0.8$, $\gamma_s+\beta_m=1.13\pm 0.05$ slightly deviates from unity.
This discrepancy is caused by the deviation of the estimated value $\sigma_s=0.89\pm 0.05$ from the theoretical value $\sigma_s=1$ \cite{cho2016hybrid}.
We think that the origin of the deviation is the crossover from HPT to continuous PT as $g \rightarrow 1$.
Finally, we confirm that the two scaling relations $\nubar_s=2\beta_s+\gamma_s$ and $\nubar_m=2\beta_m+\gamma_m$ are valid within the error ranges using the values in Table.~\ref{table1}.
However, we found that $\chi_m(t) \propto (t-t_c)^{-\gamma_m}$ is not clearly shown for $g=0.8$ again because of the crossover from HPT to continuous PT as $g \rightarrow 1$ as mentioned above.
In conclusion, we demonstrated that the scaling relations in HPT in cluster-pruning processes still hold in HPT in cluster-merging processes by using the modified $r$-ER model.

\subsection{Modified BFW model}\label{subsec:bfw}
we next study a fundamentally different model of the HPT, which does not rely on partitioning the system into sets, and identify the same features as obtained for the $r$-ER model.

To test the universality of the picture presented in \ref{subsec:rer} we study a modified version of the model proposed by Bohman et al (referred to as the $m$-BFW model)~\cite{bohman2004avoidance, chen2011explosive}, which does not contain any partitions in the system, instead the suppression effect arises in a self-organized manner via link rejection and a dynamic cap on the maximum size component allowed.

The $m$-BFW model is characterized by three parameters, denoted by $k$ the cap, $u$ the number of attempted link additions, and $h$ the lower bound on the fraction of attempted link additions that must be accepted.
While $h$ is a constant in the range$(0,1]$, $k$ and $u$ evolve in time.
The model is defined as follows:

\begin{figure*}[!htb]
    \centering
    \includegraphics[width=0.99\linewidth]{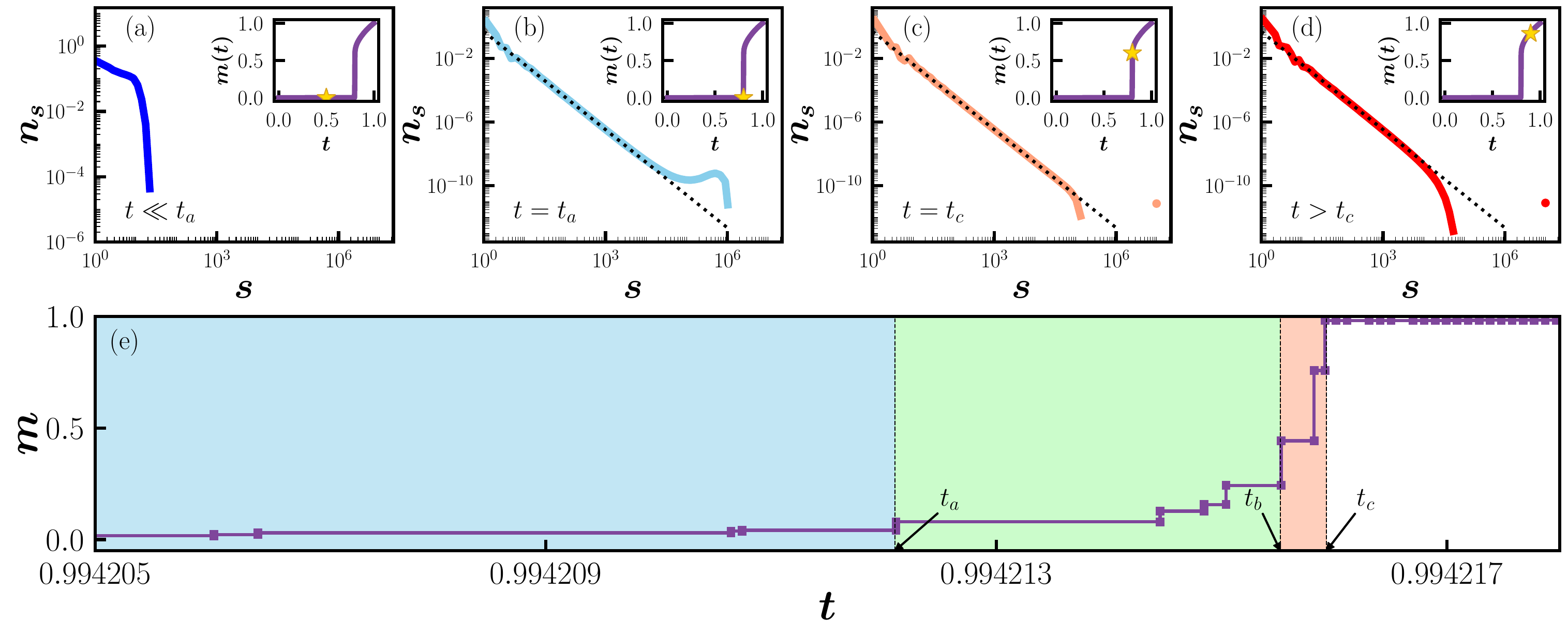}
    \caption{
        Plots of the cluster size distributions of $m$-BFW model at (a) $t \ll t_a$, (b) $t=t_a$, (c) $t=t_c$, and (d) $t > t_c$.
        Inset: Schematic plots of the order parameter $m(t)$ vs $t \equiv L/N$ with a point ($\star$) indicating the corresponding time.
        (e) Plot of the order parameter with characteristic points $\{t_a,t_b,t_c\}$ in the evolution of a single sample of the $m$-BFW model with a system size $N=1.0 \times 10^4$ for $h=0.2$.
    }
    \label{fig:fig2}
\end{figure*}

\begin{itemize}
    \item[(i)] First, the system is initialized with $N$ isolated nodes, and no links, $L=0$. Initially, $k(0)=2$ and $u(0)=0$ are set.
    \item[(ii)] Two nodes are selected randomly from two different clusters of size $s_1$ and $s_2$ and the number of trials is updated $u \to u+1$.
        A link between the selected nodes is formed and that link is added ($L\to L+1$) if
        \begin{itemize}
            \item[(ii-1)]
                $s\equiv s_1+s_2 \leq k$
            \item[(ii-2)] If $s\equiv s_1+s_2 > k$ but $L/u \leq g$, then $k \to s_1+s_2$ is set.
        \end{itemize}
        Otherwise no link is formed and $L, k$ remain unaltered.
    \item[(iii)] Return to step (ii) and repeat the loop until a single cluster of size $N$ is left.
\end{itemize}

In the original BFW model, $h(k)$ depends on $k$ as $h(k)=1/2+(2k)^{-1/2}$.
Instead here for the $m$-BFW model, $h$ is constant.
This difference yields an HPT.
For details, see Appendix~\ref{app:diff_BFW}.
In this model, $k$ controls the size of the largest cluster for a given $L$.
The number of trials to add links is counted by $u$.
Thus, $L/u$ indicates the acceptance rate of the link attachment for a given $u$, which is lower bounded by $h$.
For convenience, we use $t\equiv L/N$ as time (corresponding to the so-called event-time).
Note that for $h=1$, the $m$-BFW model is reduced to the ER model.

Large clusters in the $m$-BFW model are subject to two competing factors: suppression of growth by the model rule, and higher growth rate, proportional to their size.
Accordingly, the largest cluster grows in a boom-and-bust manner resulting in a Devil's staircase pattern.
This process effectively performs a role similar to the back-and-forth dynamics across the set boundary in the $r$-ER model.
In fact, the cluster evolution of the $m$-BFW model is similar to that of the $r$-ER model.
We find that there also exist three time intervals in the $m$-BFW model, denoted by $[0,t_a]$, $[t_a, t_b]$, and $[t_b, t_c]$.
The characteristic times $t_a$, $t_b$, and $t_c$ depend on $N$ as shown in Fig.~\ref{fig:fig2}.
In each time interval, the clusters evolve in a manner similar to that in the corresponding regime of the $r$-ER model.
In the early time regime $t \in [0, t_a]$, a bump containing a large number of medium-sized clusters forms, and its size increases in the cluster-size distribution. In the intermediate time regime $[t_a, t_b]$, the bump shrinks and the giant cluster grows rapidly. In the late time regime $[t_b, t_c]$, the bump is eroded and a pure power-law cluster size distribution is built up at $t_c$. In the late regime, the cluster merging proceeds via the ER dynamics, but with the cluster configuration $n_s(t_b)$ as an initial condition. Note that the characteristic time $t_b$ corresponds to $t_g$ of the $r$-ER model. $t_g$ is determined explicitly as $m(t_g)=1-g$ in the $r$-ER model; however, $t_b$ is determined in a self-organized manner.
\begin{figure}[!htb]
    \centering
    \includegraphics[width=0.80\linewidth]{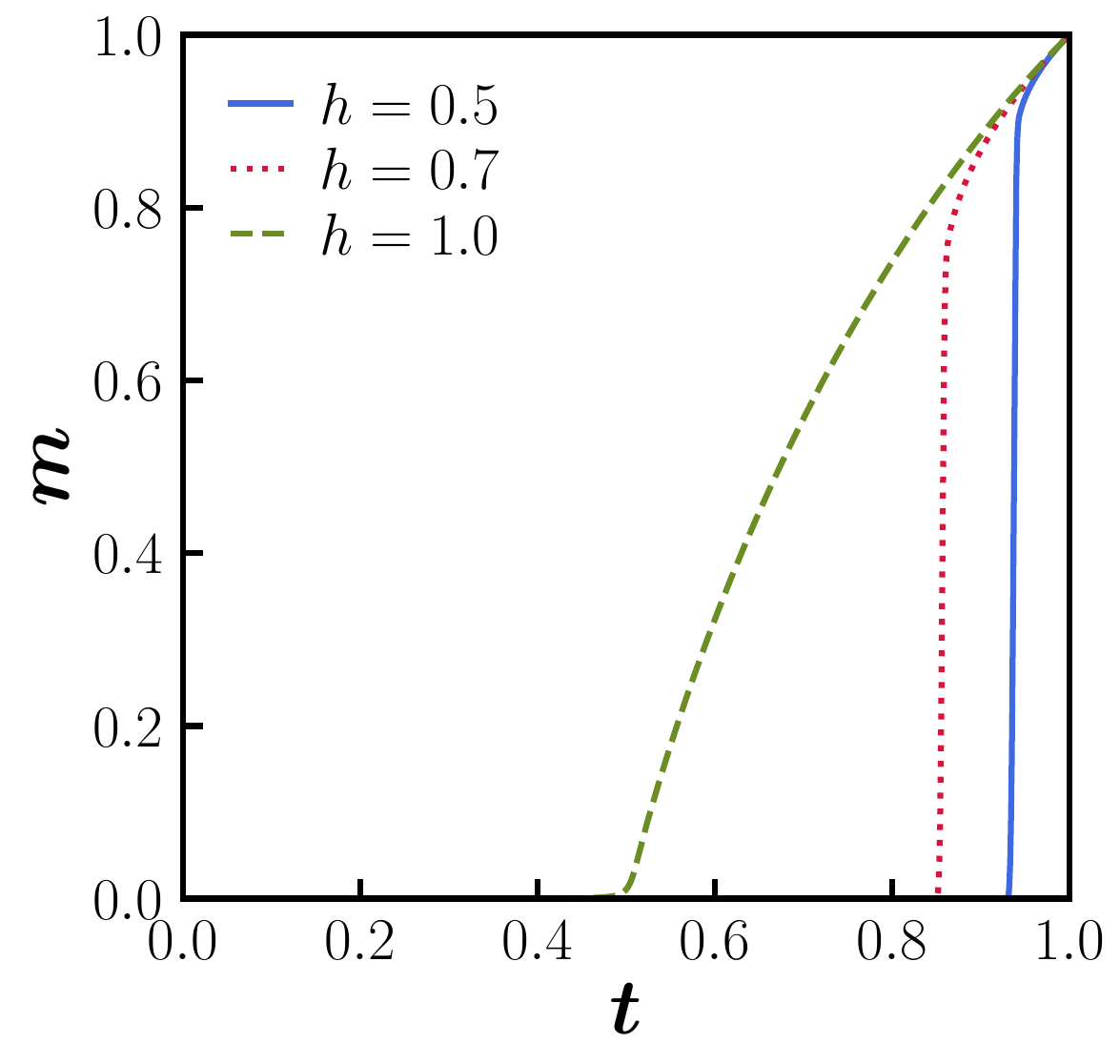}
    \caption{
        Plot of the order parameter $m(t)$ vs $t\equiv L/N$ in the $m$-BFW model for $h=1.0$ (dashed, the ER case), $0.7$ (dotted), and $0.5$ (solid).
    }   \label{fig:fig3}
\end{figure}
Fig.~\ref{fig:fig3} shows a plot of $m$ vs $t$ for different values of the model parameter $h$.
We find that the $m$-BFW model exhibits HPTs for $h < 1$, but a continuous transition for $h=1$.

\begin{figure}[!htb]
    \centering
    \includegraphics[width=0.99\linewidth]{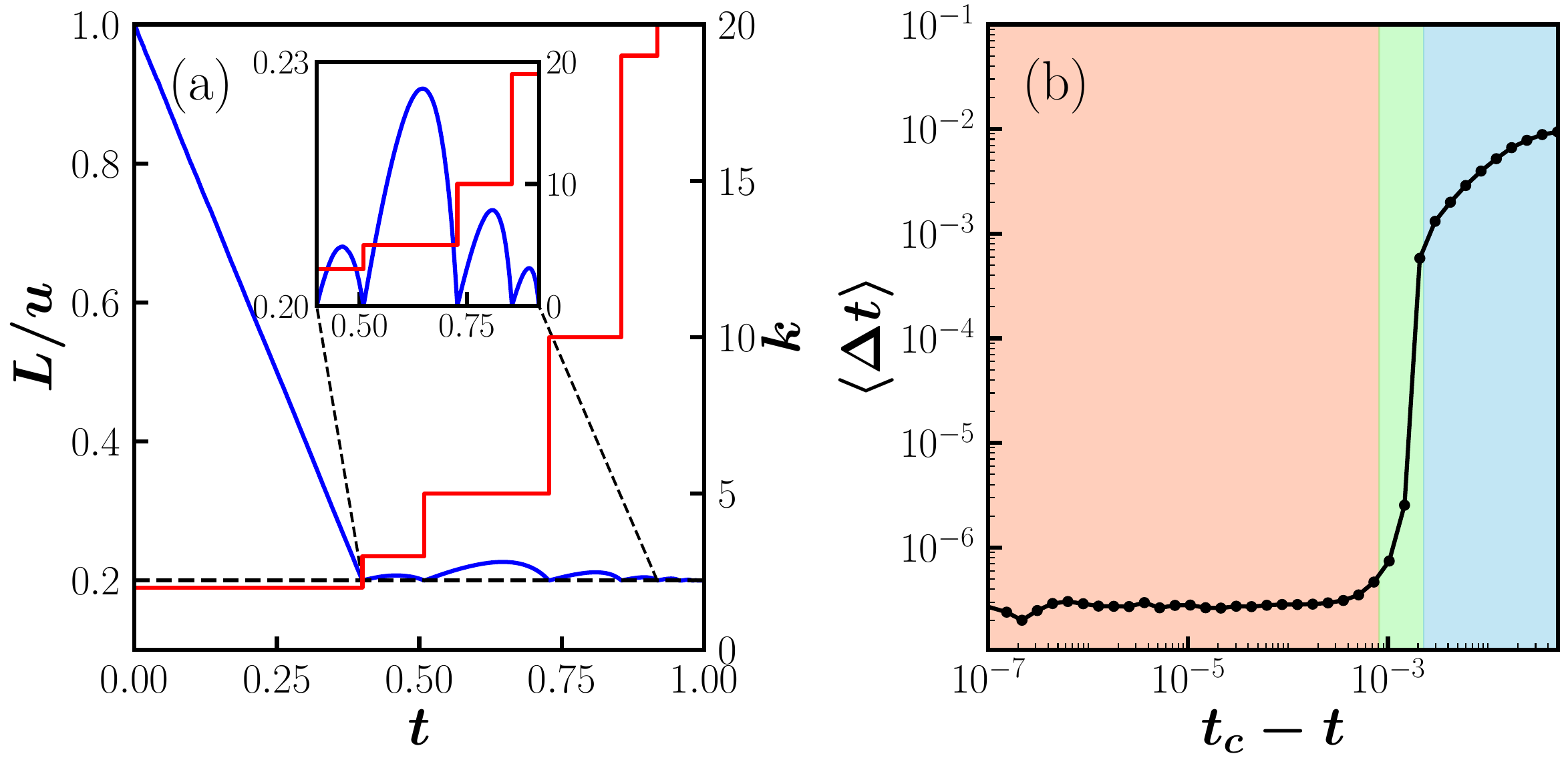}
    \caption{
        (a) Acceptance rate $L/u$ (left axis) and component size cap $k$ (right axis) vs time $t$ for $h=0.2$ in the $m$-BFW model.
        The acceptance rate $L/u$ exhibits increasing and then decreasing behavior on each plateau of $k$ as $t$ increases.
        The size cap $k$ exhibits a Devil's staircase pattern for a single configuration.
        Inset: Zoomed-in plot of the main panel in a given region.
        (b) Plot of the mean step width of the Devil's staircase $\langle \Delta t \rangle$ vs $t$ in the form of $t_c(N)-t$.
        The regimes $[0, t_a]$, $[t_a, t_b]$, and $[t_b, t_c]$ are shaded in blue, green, and red (from right to left), respectively.
        The data are obtained from $10^5$ configurations with a system size of $1.024\times 10^7$.
    } \label{fig:accept}
\end{figure}

In Fig.~\ref{fig:accept}(a), we plot the acceptance rate $L/u$ and the growth limit $k$ vs $t$. We show that $k$ increases in a Devil's staircase pattern. For a given staircase plateau, the acceptance rate $L/u$ begins to increase and then it decreases.
The first part is because $k$ is upgraded each time $L/u$ decreases to $h$, thus links can be more readily attached and the acceptance rate $L/u$ increases.
As the size of the largest cluster increases and reaches the limit $k$, the number of successful link attachment attempts declines repeatedly, which reduces $L/u$ back to $h$.
During this process, $k$ remains constant, therefore, $k$ increases following a Devil's staircase pattern.
We plot the average width of the staircases $\langle \Delta t \rangle$ over different configurations as a function of $t_c(N)-t$ in Fig.~\ref{fig:accept}(b).
$\langle \Delta t \rangle$ decreases in a power-law manner for $t < t_a(N)$, decreases rapidly in $[t_a(N), t_b(N)]$, and is almost constant in $[t_b(N), t_c(N)]$.
Thus, $t_b(N)$ can be regarded as a crossover point to the ER limit, and for $t_b(N) < t < t_c(N)$, $\langle \Delta t \rangle \approx 1/N$.
This means that almost every trial of the link attachment was accepted, particularly for giant clusters without any restrictions and the dynamics of the link attachment follows that of the ER model.

\begin{figure}[!htb]
    \centering
    \includegraphics[width=0.99\linewidth]{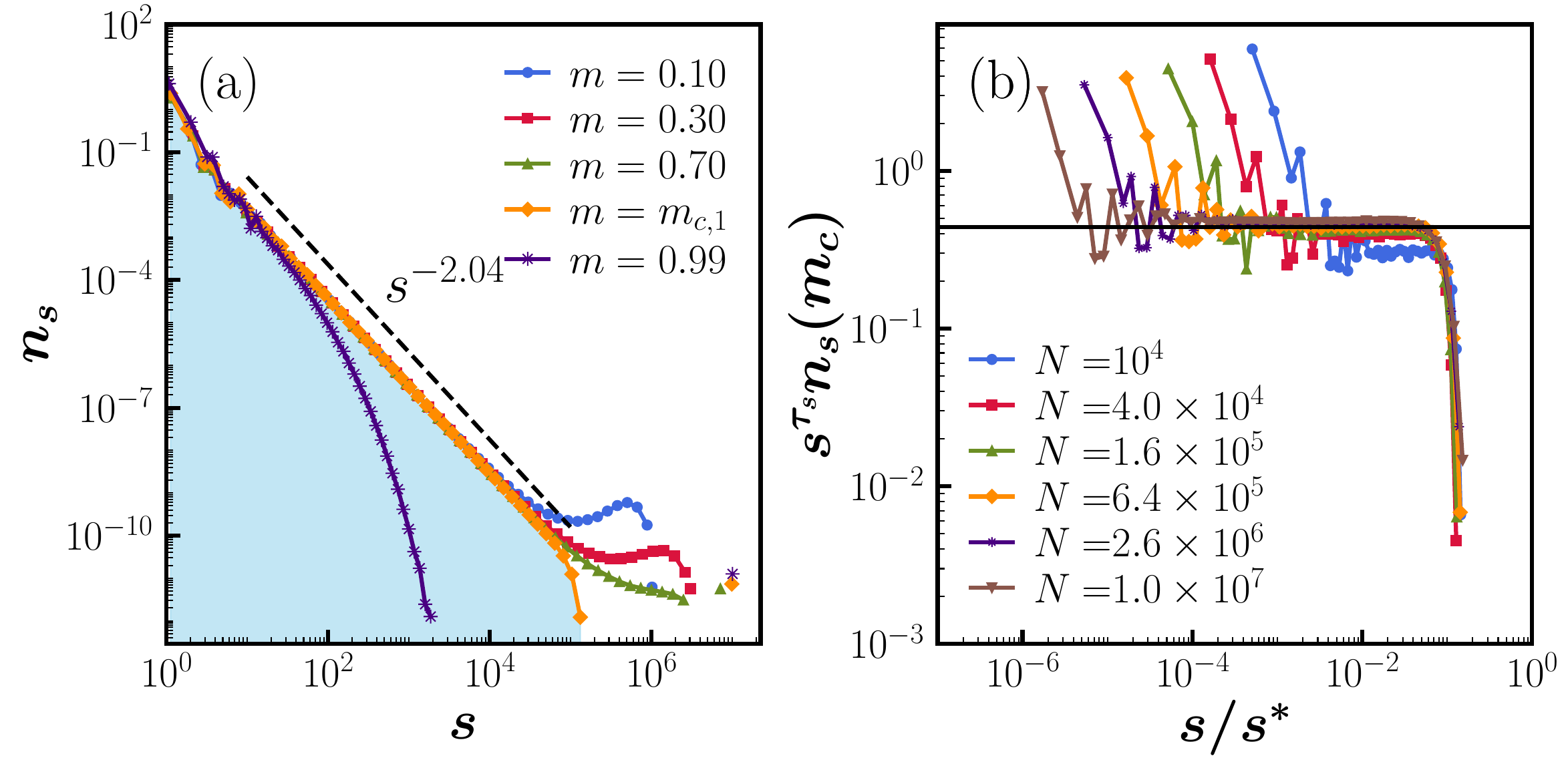}
    \caption{
    (a) Plot of the cluster-size distribution $n_s(m;N)$ vs cluster size $s$ for the $m$-BFW model with $h=0.2$.
    Data are collected from each configuration when the giant cluster size first became larger than $mN$.
    $m_{c,1}(N)$ is chosen when the bump in the cluster size distribution disappears completely, and $n_s(m_{c,1};N) \sim s^{-\tau_s}\exp{(-s/s^*)}$ with $\tau_s\approx 2.04$.
    $s^*$ is the characteristic size due to the finite-size effect.
    The area $\sum_s^\prime n_s(m;N)$ (colored in blue (gray)) is denoted by $1-m_{c,2}(N)$, which indicates the fraction of nodes contained in all finite clusters.
    Accordingly, the order parameter was obtained as $m_{c,2}(N)$.
    Simulations were performed in systems of size $N=1.024\times 10^7$ with the model parameter $h=0.2$.
    Data are averaged over $10^5$ configurations.
    (b) Plot of the finite-size scaling analysis.
    The characteristic size $s^*\sim N^{1/{\sigma_s\nubar_s}}$ with $\sigma_s \nubar_s\approx 1.20.$
    }   \label{fig:n_s}
\end{figure}

Due to the complexity of the transition, finite size scaling should be studied with special care.
We determine a transition point $t_c(N)$ as follows: We denote the sample index by $i$.
At the first time when the largest cluster size is equal to or larger than $m$ the size distribution of finite clusters divided by $N$ for sample $i$ is $n_s^\isample(m,N)$.
We take an average over all samples:
\begin{equation}
    n_s(m;N)\equiv \langle n_s^\isample(m;N)\rangle_i.
    \label{eq:n_s}
\end{equation}

\begin{figure}[!hb]
    \centering
    \includegraphics[width=0.99\linewidth]{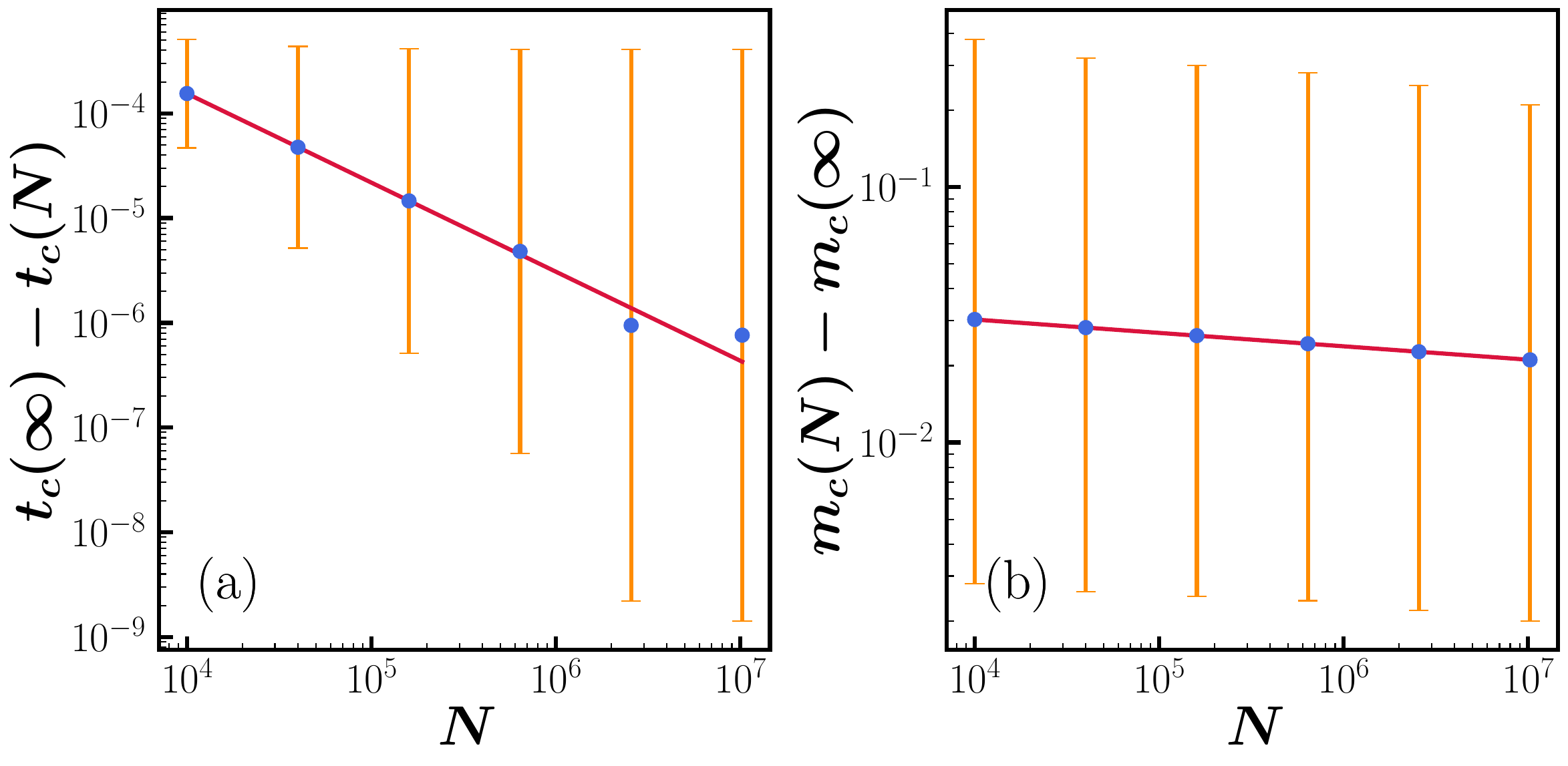}
    \caption{
    Finite size scaling plots for the m-BFW model with $h=0.2$.
    (a) $t_c(\infty)-t_c(N)$ vs $N$ with error bars for $t < t_c(\infty)$.
    It behaves as $\sim N^{-1/\nubar_m} = N^{-0.85}$.
    (b) Plot of $m_c(N)-m_c(\infty)$ vs $N$, where $m_c(N) \equiv \langle m_c^{(i)}(N) \rangle_i$.
    By using the relation $m_c(N)-m_c(\infty) \sim N^{-\beta_m/\nubar_m}$, $\beta_m/\nubar_m \approx 0.05$ for $h=0.2$.
    The error bars represent the standard deviation of $m_c(N)$.
    }  \label{fig:t_c}
\end{figure}

When $m$ is small, the size distribution $n_s(m;N)$ of finite clusters contains a bump.
However, there exists a characteristic value $m_{c,1}(N)$ at which the bump disappears completely, and the distribution $n_s(m_{c,1};N)$ exhibits power-law behavior.
Therefore, $n_s(m_{c,1};N)$ can be expressed as $\sim s^{-\tau_s}\exp(-s/s^*)$, as shown in Fig.~\ref{fig:n_s}(a).
We directly measured the exponent $\tau_s$ as $\tau_s=2.04\pm 0.02$ for $h=0.2$.
The characteristic cluster size $s^*$ behaves as $(t_c-t)^{-1/\sigma_s}$ for $t < t_c$ and $\sim N^{1/\sigma_s\nubar_s}$ at $t_c$.
We estimate that $1/\sigma_s\nubar_s\approx 0.83$ ($\sigma_s\nubar_s\approx 1.20$), as shown in Fig.~\ref{fig:n_s}(b).
We define $m_{c,2}(N)\equiv 1-\sum^\prime_s sn_s(m_{c,1};N)$, where the prime denotes summation over finite clusters.
Next, for each sample $i$, we find $t_c^\isample(N)$, where the giant cluster size is increased to $m_c^{(i)}(N) \ge m_{c,2}$ for the first time.
Subsequently, $t_c(N)$ is determined as $\langle t_c^{(i)}(N) \rangle_i$, and $m_c(N)$ is determined as $\langle m_c^{(i)}(N)\rangle_i$.
Based on $t_c(N)$, we determine $t_c(\infty)$ using the finite-size scaling theory, and the plot of  $t_c(\infty)-t_c(N)$ vs $N$ is assumed to follow a power law as $\sim N^{-1/\nubar_m}$ by choosing $t_c(\infty)$~(see Fig.~\ref{fig:t_c}(a)).
We estimate $t_c(\infty)=0.9953$ and $1/\nubar_m \approx 0.85$ ($\nubar_m \approx 1.18$) for $h=0.2$.
We also determine $m_c(\infty)$ using finite-size scaling: $m_c(N)-m_c(\infty)\sim N^{-\beta_m/\nubar_m}$ by choosing the proper $m_c(\infty)$ showing a power-law decay~(see Fig.~\ref{fig:t_c}(b)).
We obtain $m_c(\infty)\approx 0.9555$ and $\beta_m/\nubar_m = 0.05$ for $h=0.2$.
Therefore, $\beta_m \approx 0.062$ is estimated.
We also directly measured the exponent $\beta_m$ as $0.06\pm 0.01$ (see Fig.~\ref{fig:abfss}).

The value of $t_c(N)$ is determined as the mean $\langle t_c^{(i)}(N)\rangle_i$ over the different samples; it has a relatively large error bar.
The standard deviation of $\{t_c^\isample(N)\}$ is defined as:
\begin{align}
    \sigma_c^2(t_c(N))\equiv \frac{1}{N_{\rm config}}\sum_{i=1}^{N_{\rm config}} (t_c^\isample(N)-t_c(N))^2. \quad
\end{align}

\begin{figure*}[!htb]
    \centering
    \includegraphics[width=0.99\linewidth]{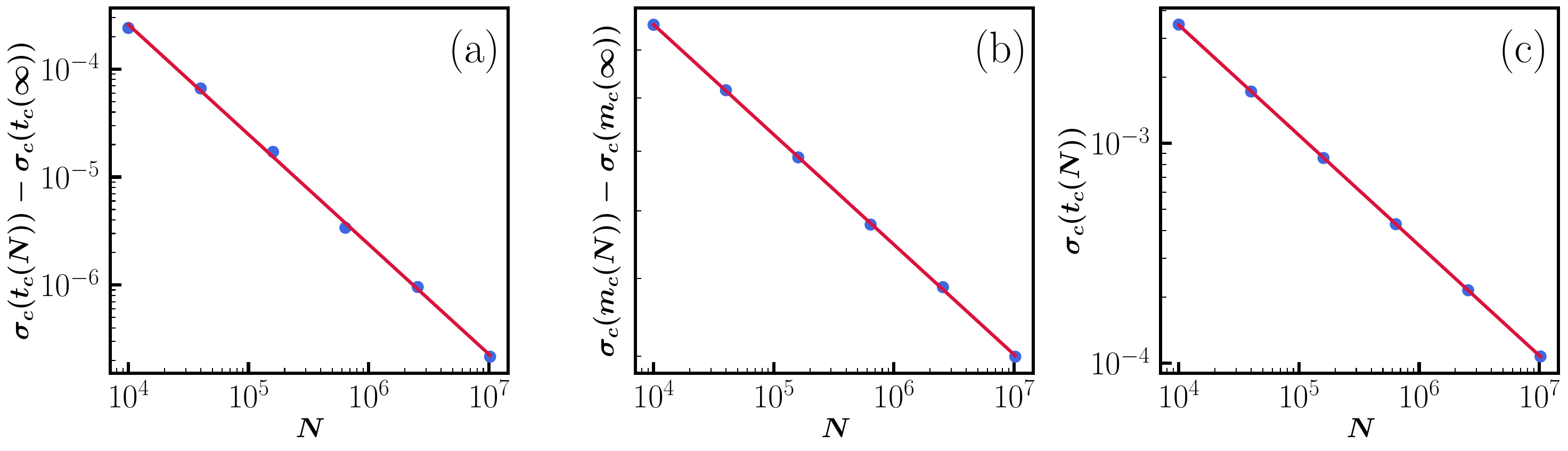}
    \caption{
        (a) Plot of the finite size scaling behavior of $\sigma_c(t_c(N))-\sigma_c(t_c(\infty))$ vs $N$ for the $m$-BFW model with $h=0.2$.
        Here $\sigma_c(t_c(\infty)) \approx 9.4\times 10^{-4}$ is chosen to obtain good scaling.
        The finite size scaling exponent is found to be $1.02$.
        (b) Plot of the standard deviation $\sigma_c(m_c(N))-\sigma_c(m_c(\infty))$ of $m_c(N)$ with $\sigma_c(m_c(\infty))\approx 0.0036$.
        The finite-size scaling exponent is found to be 0.09.
        (c) Similar plot for the $r$-ER model with $g=0.5$: $\sigma_c(t_c(N))$ decays in a power-law manner as $\sim N^{-0.5}$.
        Therefore, $\sigma_c(t_c(\infty))=0$.
    }\label{fig:std}
\end{figure*}

\begin{figure}[!htb]
    \centering
    \includegraphics[width=0.99\linewidth]{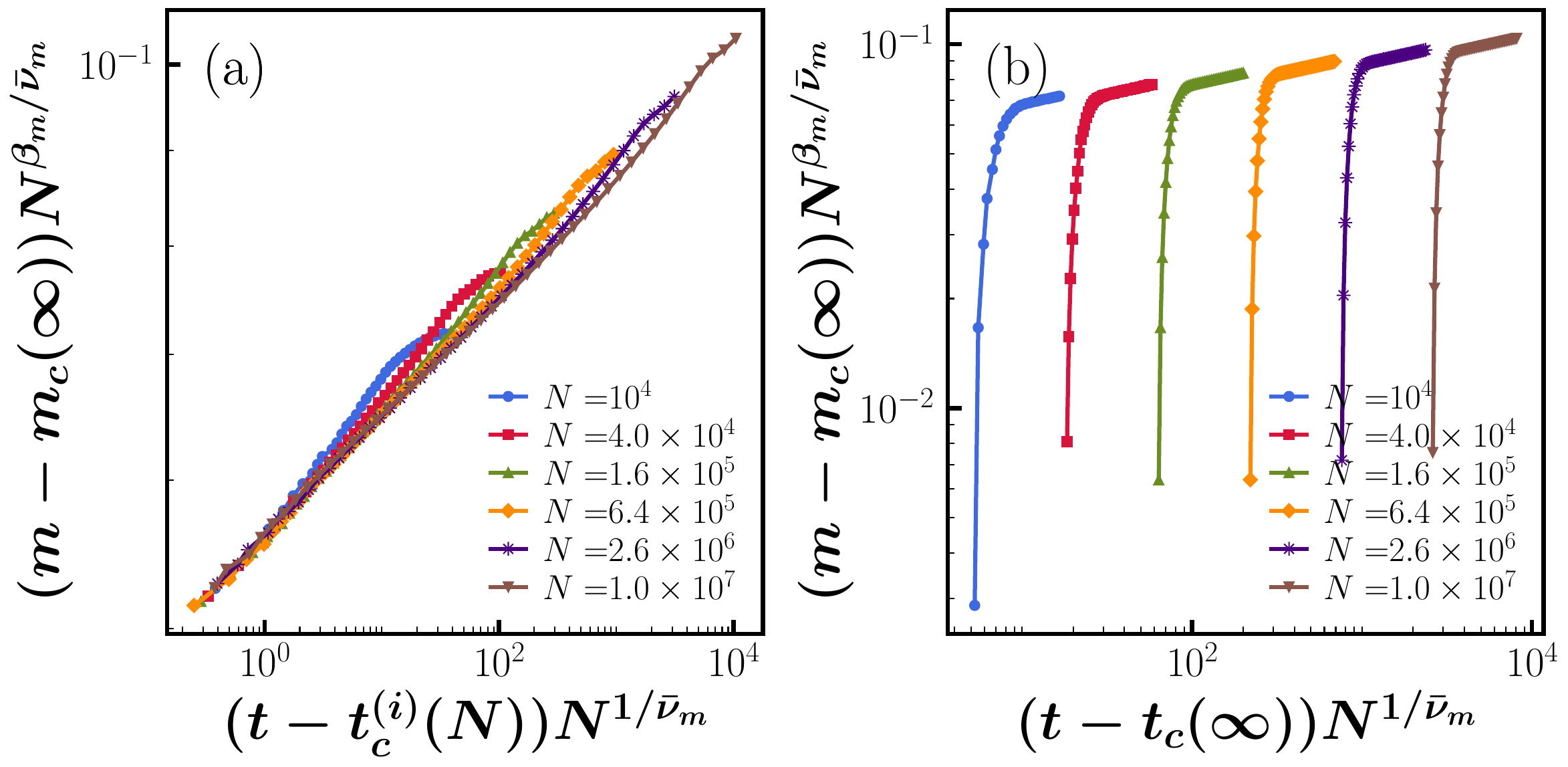}
    \caption{
        (a) Finite-size scaling of order parameter $m$ for $h=0.2$ using $\bart(N)$.
        By taking $\beta_m=0.059$ and $\nubar_m=1.124$, the data were reasonably well collapsed onto a single curve.
        (b) Finite-size scaling of order parameter $m$ in the conventional method using $t-t_c(\infty)$.
        The data did not collapse onto a single curve.
    }   \label{fig:abfss}
\end{figure}

We find that as $N\to \infty$, $\sigma_c(t_c(\infty))$ does not reduce to zero, but converges to a finite value.
The difference scales as $\sigma_c(t_c(N))-\sigma_c(t_c(\infty)) \sim N^{-1.02}$, as shown in Fig.~\ref{fig:std}(a).
This corresponds to $\sigma_c(m_c(N))-\sigma(m_c(\infty))\sim N^{-0.09}$ with a non-zero $\sigma_c(m_c(\infty))$.
However, for the $r$-ER model, $\sigma_c(t_c(\infty))=0$ and $\sigma_c(t_c(N)) \sim N^{-0.5}$, as shown in Fig.~\ref{fig:std}(b).

Owing to the large error bars of $t_c(N)$ and $\sigma_c(t_c(\infty))\ne 0$, the conventional finite-size scaling theory in the region $t> t_c(\infty)$ fails.
Hence, we construct a novel framework for finite-size scaling analysis in the supercritical region as follows.
We plot $(m-m_c(\infty))N^{\beta_m/\nubar_m}$ vs $ (t-t_c^\isample(N))N^{1/\nubar_m}$ for different samples and system sizes in double logarithm scales in Fig.~\ref{fig:abfss} (a).
By choosing $\beta_m=0.059$ and $\nubar_m=1.124$, we find that the data collapse reasonably well onto a single curve.
Note that the plot is based on
\begin{equation}
    \bart(N) \equiv t-t_c^\isample(N)   \label{eq:aaa}
\end{equation}
meaning that the transition points of each sample for a given $N$ are adjusted to a single point instead of using their mean value $t_c(N)$.
Note that in the conventional finite-size scaling theory, the scaling plot is drawn as $(m-m_c(\infty))N^{\beta_m/\nubar_m}$ vs $(t-t_c(\infty))N^{1/\nubar_m}$.
From this plot, we find that the data from different system sizes are not collapsed onto a single curve, as shown in Fig.~\ref{fig:abfss}(b).
Hence, the conventional finite-size scaling approach established for the second-order percolation transition does not work well for the $m$-BFW model.

\begin{figure}[!htb]
    \centering
    \includegraphics[width=0.99\linewidth]{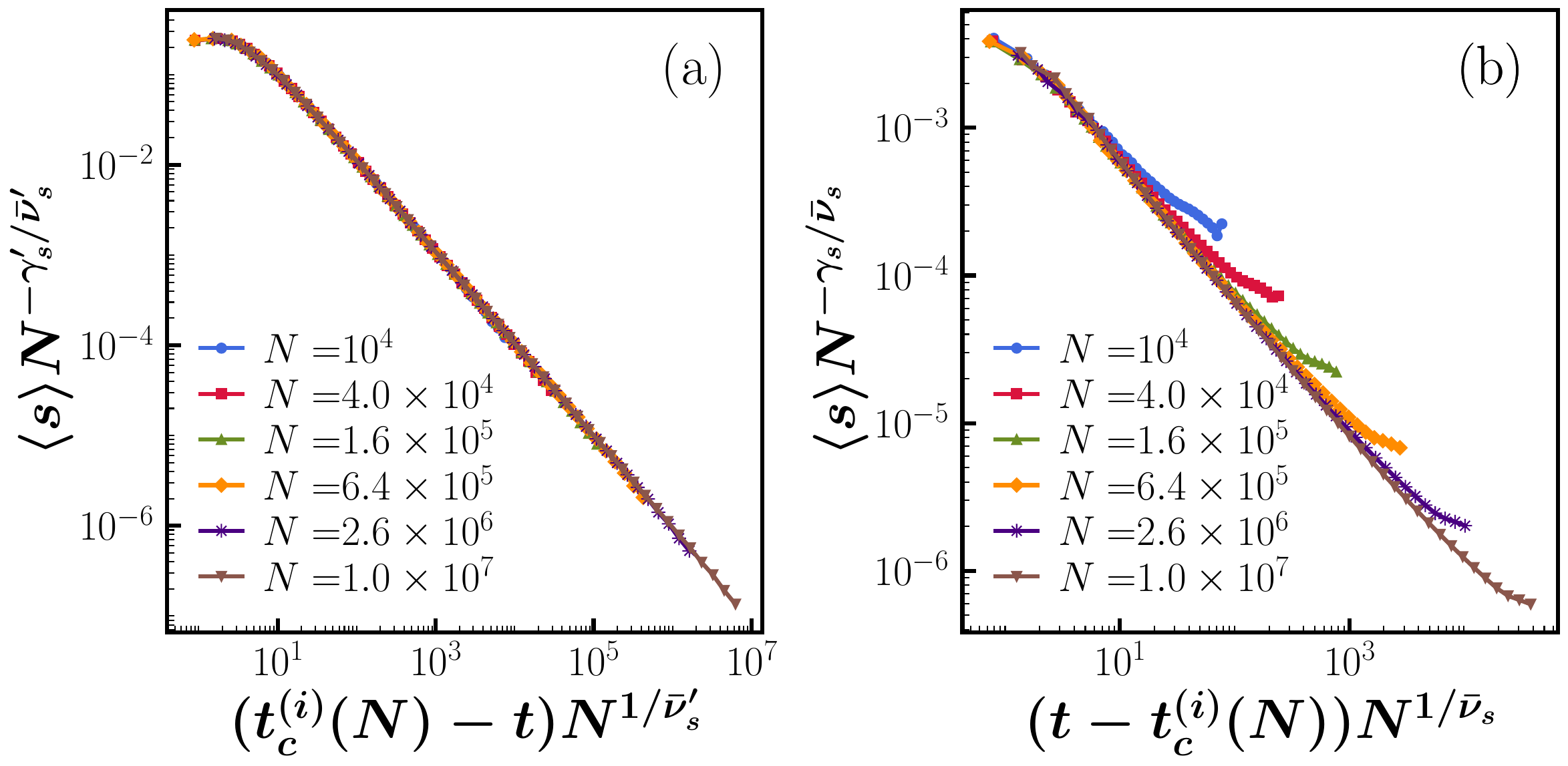}
    \caption{
        (a) Scaling plot of mean cluster size $\langle s \rangle N^{-\gamma_s^\prime/\nubar_s^\prime}$ vs $(t_c^\isample(N)-t)N^{1/\nubar_s^\prime}$ for different system sizes (a) for $t < t_c(\infty)$ (in the subcritical regime) using $\gamma_s^\prime=1.015$, $\nubar_s^\prime=1.020$ and
        (b) for $t > t_c(\infty)$ (in the supercritical regime) using $\gamma_s=0.918$ and $\nubar_s=1.031$.
    }   \label{fig:gamma_s}
\end{figure}

We investigate the mean cluster size of finite clusters for systems of size $N$.
For each configuration $i$, we obtain the cluster size distribution $n_s^\isample(t)$ and then obtain the mean cluster size of the configuration $i$, $\bar s^\isample(t)\equiv \sum^\prime_s s^2 n_s^\isample(t)/\sum^\prime_s s n_s^\isample(t)$.
Because the transition point $t_c^\isample(N)$ of the configuration $i$ fluctuates significantly, we first plot the mean cluster size in scaling form as $\bar s^\isample(t) N^{-\gamma_s^\prime/\nubar_s^\prime}$ vs $(t_c^\isample(N)-t)N^{1/\nubar_s^\prime}$.
Next, we take average of configuration index $i$, and obtain the plot shown in Fig.~\ref{fig:gamma_s} (a) and (b), where $\langle s \rangle \equiv \sum_i{\bar s}^\isample/N_s$.
$N_s$ represents the number of configurations.
In Fig.~\ref{fig:gamma_s} (a) and (b), the scaling plot $\langle s(t) \rangle N^{-\gamma_s/\nubar_s}$ vs $|\bart(N)| N^{1/\nubar_s}$ exhibits data collapse behavior for different samples and system sizes, with the choice of $\gamma_s^\prime=1.015$ and $\nubar_s^\prime=1.020$ for $t < t_c$, and $\gamma_s=0.918$ and $\nubar_s=1.031$ for $t > t_c$ for $h=0.2$.

\begin{figure}[!htb]
    \centering
    \includegraphics[width=0.99\linewidth]{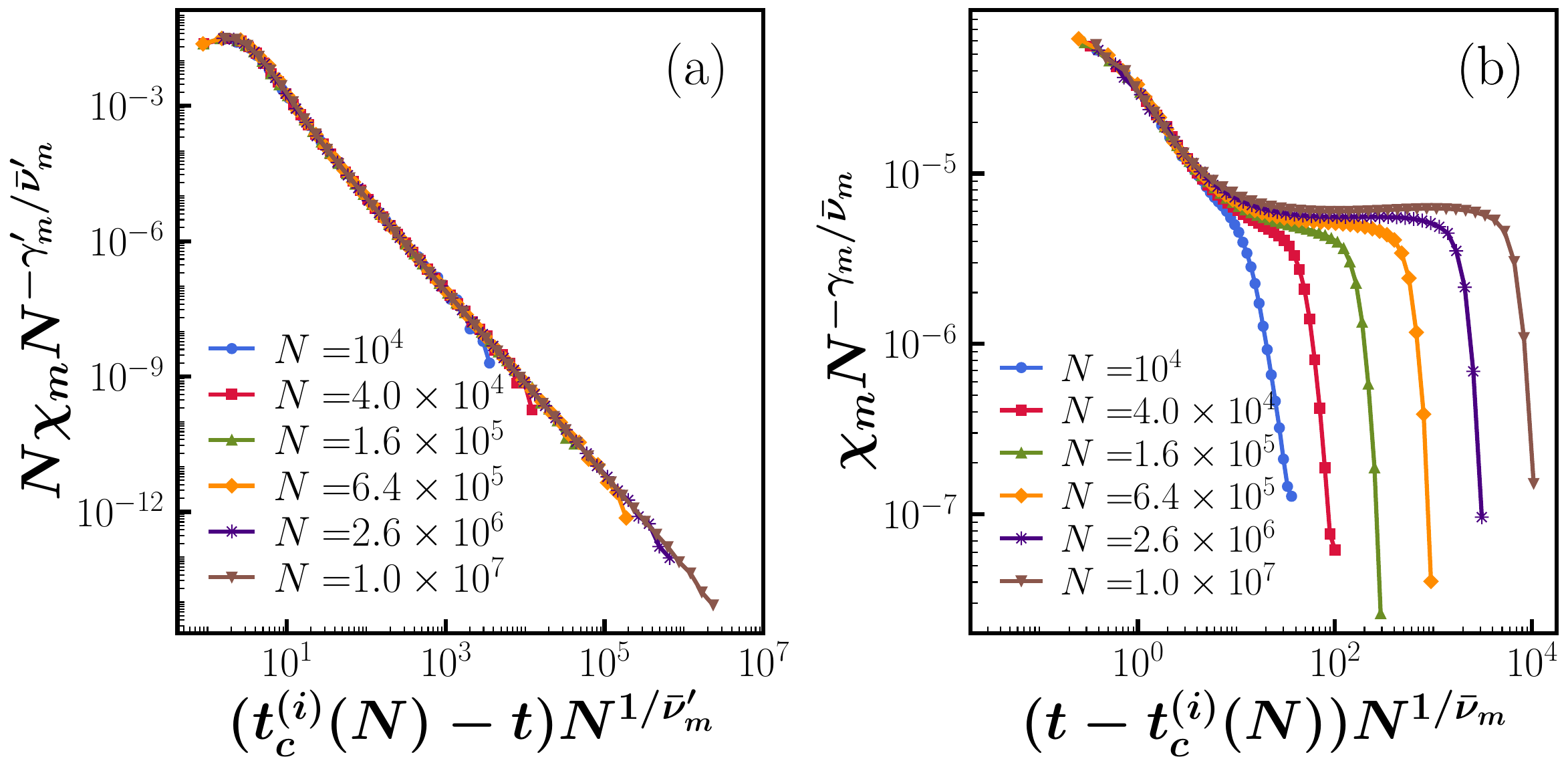}
    \caption{
        (a) Scaling plot of susceptibility $\chi_m N^{-\gamma_m^\prime/\nubar_m^\prime}$ vs $(t_c^\isample(N)-t)N^{1/\nubar_m^\prime}$ for different system sizes (a) for $t < t_c(\infty)$ (in the subcritical regime) using $\gamma_m^\prime=2.041$, $\nubar_m^\prime=1.020$ and
        (b) for $t > t_c(\infty)$ (in the supercritical regime) using $\gamma_m=1.034$ and $\nubar_m=1.124$
    }   \label{fig:gamma_m}
\end{figure}

Next, we examine the susceptibility representing the fluctuations of the largest cluster $N\chi_m$ for $t< t_c$ and of the order parameter for $t > t_c$, formulated as $\chi_m=N(\langle m^2 \rangle - \langle m \rangle^2)$.
In Fig.~\ref{fig:gamma_m}(a) and (b), we attempt to show the data-collapse behavior in the scaling plot of $\chi_m(t) N^{-\gamma_m/\nubar_m}$ vs $|\bart(N)| N^{1/\nubar_m}$ with the choices of $\gamma_m^\prime=1.020$ and $\nubar_m^\prime=1.020$ for $t < t_c$, and $\gamma_m=1.034$ and $\nubar_m\approx 1.124$ for $t > t_c$ and $h=0.2$.
However, the data-collapse is not satisfactory, particularly somewhat in broad range of $t> t_c$ near $t_c^+$.
Therefore, the critical exponent $\gamma_m$ for $t > t_c$ is not well defined.
The numerical values of the critical exponents for other $h$ parameters are listed in Table.~\ref{table2}.

We have verified the scaling relations between the two sets of critical exponents for $t > t_c$: one set is $\{\tau_s, \sigma_s, \beta_s, \gamma_s, \nubar_s\}$, and the other set is $\{\beta_m, \gamma_m, \nubar_m\}$.
We first check the scaling relations among the sets associated with the cluster size distribution.
Using the directly measured values $\tau_s=2.04 \pm 0.02$ in Fig.~\ref{fig:n_s} and $\gamma_s=0.87 \pm 0.07$ in Fig.~\ref{fig:gamma_s}, we estimate $\sigma_s = (3-\tau_s)/\gamma_s = 1.11 \pm 0.11$, $\beta_s = (\tau_s-2)/\sigma_s = 0.04 \pm 0.02$, and $\nubar_s = 2\beta_s+\gamma_s = 0.96 \pm 0.08$.
We obtained $\nubar_s\approx 1.031$ as a direct estimate using the data-collapse method in Fig.~\ref{fig:gamma_s}.
Thus, the two values of $\nubar_s$ are in agreement within the error range.

Next, we confirm that for $t> t_c$, $\sigma_s\approx 1$.
This is consistent with the theoretical prediction $\sigma_s=1$, resulting from the argument that the size distribution of finite clusters is analytic~\cite{costa,cho2016hybrid}(see Appendix~\ref{app:deriv_sigma}).

\begin{table*}[!htb]
    \renewcommand{\arraystretch}{1.2}
    \centering
    \setlength{\tabcolsep}{1.5em}
    \begin{tabular}{*{7}{c}}
        \hline\hline
        $h$ & $\tau_s$        & $\sigma_s$      & $\beta_s$       & $\nubar_s$        & $\gamma_s$      & $\gamma_s^\prime$ \\
        \hline
        0.2 & $2.04 \pm 0.02$ & $1.11 \pm 0.11$ & $0.04 \pm 0.02$ & $0.96 \pm 0.08$   & $0.87 \pm 0.07$ & $1.02 \pm 0.02$   \\
        0.5 & $2.21 \pm 0.03$ & $1.04 \pm 0.08$ & $0.21 \pm 0.05$ & $1.26 \pm 0.19$   & $0.76 \pm 0.03$ & $1.08 \pm 0.06$   \\
        0.8 & $2.30 \pm 0.05$ & $1.02 \pm 0.11$ & $0.30 \pm 0.08$ & $1.33 \pm 0.24$   & $0.69 \pm 0.03$ & $1.15 \pm 0.03$   \\
        \hline\hline
        $h$ & $\beta_m$       & $\nubar_m$      & $\gamma_m$      & $\gamma_m^\prime$ & $\zeta$         & $\zeta^\prime$    \\
        \hline
        0.2 & $0.06 \pm 0.01$ & $1.18 \pm 0.06$ & $1.05 \pm 0.07$ & $2.08 \pm 0.07$   & $1.01 \pm 0.01$ & $1.99 \pm 0.08$   \\
        0.5 & $0.24 \pm 0.01$ & $1.42 \pm 0.11$ & $0.93 \pm 0.12$ & $2.16 \pm 0.02$   & $1.02 \pm 0.04$ & $1.79 \pm 0.03$   \\
        0.8 & $0.33 \pm 0.02$ & $1.51 \pm 0.13$ & $0.85 \pm 0.16$ & $2.36 \pm 0.05$   & $1.09 \pm 0.06$ & $1.74 \pm 0.08$   \\
        \hline\hline
    \end{tabular}
    \caption{
        List of estimated exponent values for different values of $h$.
        $t_c$ is the transition point in the thermodynamic limit estimated using finite-size scaling (FSS) analysis.
        $\tau_s$ is the exponent of the cluster size distribution at $t_c$.
        $\beta_m$ is the critical exponent of the order parameter estimated using FSS analysis.
        $\zeta$ and $\zeta^\prime$ are the exponents of the inter-cluster coalescence time distribution $P_a(z)$ estimated in the regimes $[0, t_a]$ and $[t_a, t_c]$, respectively.
    }   \label{table2}
\end{table*}

A scaling relation among the exponents $\{\beta_m, \gamma_m, \nubar_m\}$; $\nubar_m=2\beta_m + \gamma_m$ can also be verified.
We estimate $\beta_m=0.06 \pm 0.01$ in Fig.~\ref{fig:abfss}(a), $\gamma_m=1.00 \pm 0.10$ in Fig.~\ref{fig:gamma_m}.
Thus, we obtain $\nubar_m = 1.12 \pm 0.12$ from the above scaling relation.
This value is in agreement with the directly measured value $\nubar_m \approx 1.124$ using the data-collapse method in both Fig.~\ref{fig:abfss}(a) and Fig.~\ref{fig:gamma_m}(b).

Finally, we check whether the scaling relation Eq.~\eqref{eq:gamma_a_beta_m}, i.e., $\gamma_s+\beta_m=1$ holds.
The origin of the scaling relation is similar to the case of HPT with cluster pruning processes: The avalanche size there is replaced in cluster merging processes by the mean cluster size.
This relation bridges the two sets of critical exponents.
Using the values $\gamma_s=0.87\pm 0.07$ and $\beta_m=0.06\pm 0.01$, we confirm that the scaling relation $\gamma_s+\beta_m=1$ holds within the numerical accuracy.

As stated in subsection~\ref{subsec:rer} for the $r$-ER model, the inter-event time (denoted as $z_i$) is defined as the time taken for a node to move from one set to the other set.
The inter-event time distribution $P_I(z)$ exhibits a power-law decay with exponent $\zeta$ in the interval $[0,t_a]$ and exponent $\zeta^\prime$ in the interval $[t_a,t_g]$.
This inter-event time can be defined because the $r$-ER system is composed of two partitions.
However, for the $m$-BFW model, in which the system does not contain any partitions, the inter-event time cannot be defined.
Here, we use the age distribution as an alternative.
The age (also denoted as $z$) is defined as the duration of the time during which the cluster of a given node is not linked with any other cluster.
For the $r$-ER model, the age distribution $P_a(z)$ behaves similarly to $P_I(z)$: $P_a(z)\sim z^{-\zeta}$ for $t \in [0,t_a]$ and $\sim z^{-\zeta^\prime}$ for $t \in [t_a, t_g]$.
We have found for the $r$-ER model that~\cite{park2019interevent} $\zeta \approx 1.04$ and $\zeta^\prime \approx 1.87$ for $g=0.2$.
Recall that in $t \in [t_a, t_g]$, $s^*$ is constant for $t \to t_c(\infty)$ and the scaling exponent of the characteristic size of finite clusters $\sigma_s^\prime=0$.
Hence, the relation $\zeta^\prime=4-(\tau_s+\sigma_s^\prime)$ is reduced into $\zeta^\prime=4-\tau_s$ in $[t_a, t_g]$.

\begin{figure}[!htb]
    \centering
    \includegraphics[width=0.99\linewidth]{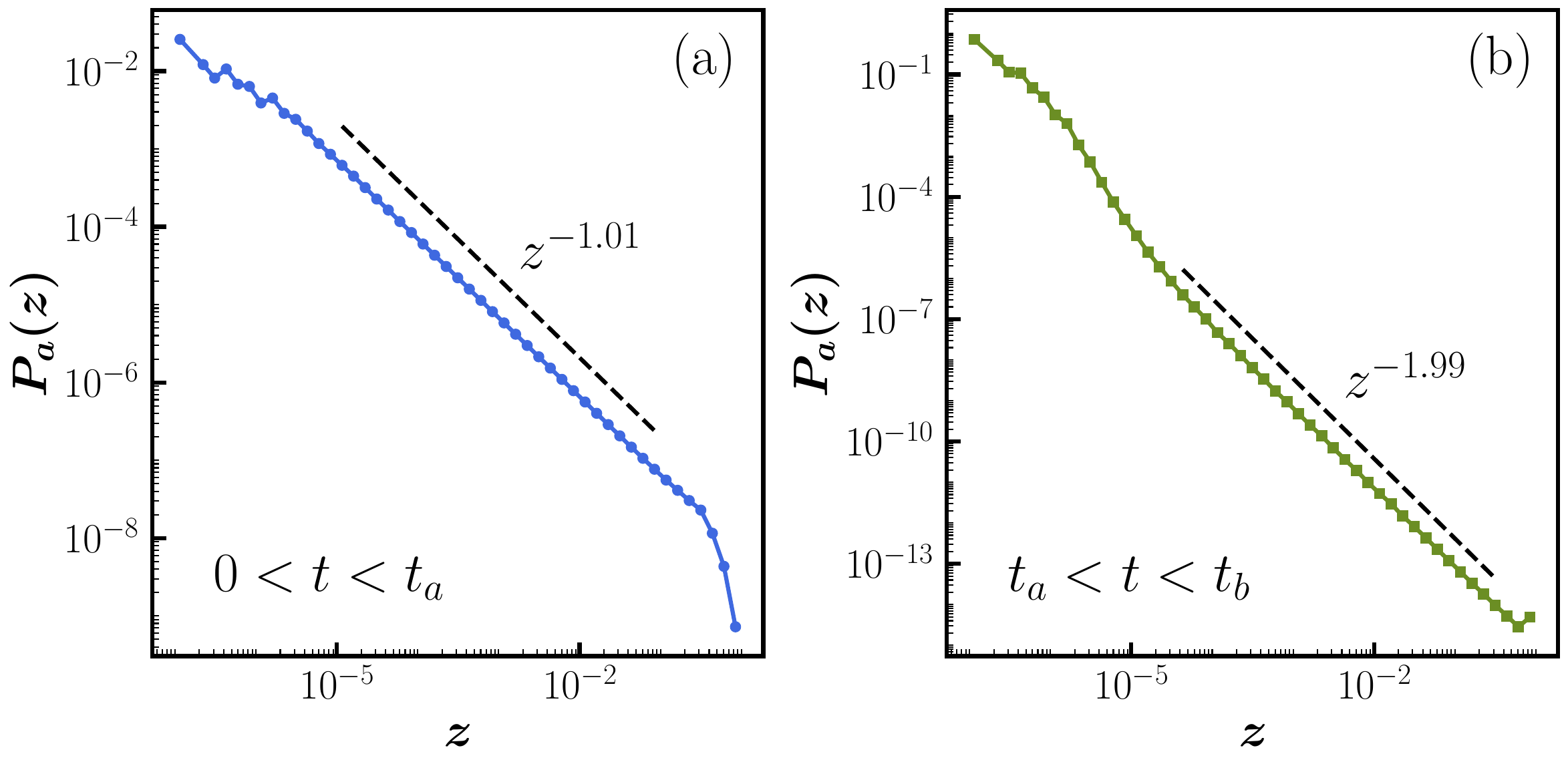}
    \caption{
        Plot of the age distributions $P_a(z)$ in the intervals (a) $t \in [0,t_a]$ and (b) $t\in [t_a, t_b]$.
        The data are obtained from $10^5$ configurations of system size $1.024\times 10^7$ for $h=0.2$}
    \label{fig:age}
\end{figure}

This result is also valid for the $m$-BFW model.
Similar to the $r$-ER model, characteristic size of the finite clusters is constant at a very short time interval $[t_a, t_b]$.
Therefore, $\sigma_s^\prime=0$ at $[t_a, t_b]$ and the relation $\zeta^\prime=4-\tau_s$ also holds.
Based on the value, $\tau_s=2.04 \pm 0.02$ we predict that $\zeta^\prime=1.96 \pm 0.02$ for $h=0.2$, which is in good agreement with the value that was directly measured, $1.99 \pm 0.08$.
Furthermore, using $\zeta \approx 1.01 \pm 0.01$ shown in Fig.~\ref{fig:age} and $\tau_s\approx 2.04\pm 0.02$, we obtain $\sigma_s^\prime=0.95 \pm 0.03$ from the scaling relation $\zeta=4-(\tau_s+\sigma_s^\prime)$ in the interval $[0,t_a]$.

\section{Conclusions}  \label{sec:conclusion}
In this paper, we considered two models, called the modified $r$-ER model and the $m$-BFW model, that exhibit hybrid percolation transitions (HPTs) in cluster merging dynamics.
Even though the dynamic rules of the cluster merging process are different for the two models, we showed that their underlying mechanisms are universal.
An HPT with cluster merging occurs following three regimes:
In the early regime of $t\in [0, t_a]$, medium-sized clusters accumulate and create a bump in cluster size distribution due to the suppression of the growth of a giant cluster. This is consistent with a ``powder keg" \cite{Friedman2009Construction}.
Here, $t_a$ is defined as the time at which the bump size reaches its maximum.
In the intermediate regime $t \in [t_a, t_g]$ for the modified $r$-ER model and $t \in [t_a, t_b]$ for the $m$-BFW model, the medium-sized clusters merge and a giant cluster rapidly increases.
Time $t_g$ or $t_b$ is a crossover point at which the suppression effect disappears. This crossover point is determined by the dynamic rule itself in the modified $r$-ER model and in a self-organized manner in $m$-BFW model.
In the final regime $t \in [t_g, t_c]$ or $t\in [t_b, t_c]$, the cluster-merging kinetics becomes effectively the ER dynamics but with the initial configuration of cluster size distribution at $t_g$ or $t_b$ determined by the evolutionary dynamics.
At $t_c$, the size distribution of finite clusters follows a power law and the system exhibits critical behavior.

The critical behavior of the HPT is characterized by two sets of critical exponents.
One set comprising $\{\beta_m, \gamma_m, \nubar_m \}$ is associated with a giant cluster, and the other set comprising $\{\tau_s, \sigma_s, \beta_s, \gamma_s, \nubar_s\}$ is associated with the size distribution of finite clusters.
Whereas the two sets of critical exponents are reduced to one set in the second-order, e.g., Bernoulli percolation transition, they remain different for HPT, for example, $\gamma_m \ne \gamma_s$. Unfortunately, due to the relatively large errors, the two sets of exponents are numerically not that different.
As the model parameter $g$ and $h$ of the modified $r$-ER and $m$-BFW models, respectively, change, the values of critical exponents continuously vary.
However, the exponents in each set follow scaling relations respectively and these two sets are not independent, connected by a scaling relation $\gamma_s=1-\beta_m$.
The relation is the counterpart of that for HPT with the cluster pruning process, with the critical exponents associated with avalanche size distribution (cluster pruning) replaced by the exponents associated with the cluster size distribution (cluster merging). This way a unified picture for HPT is established.

The critical exponents are also related to the distribution of age, defined as the duration of time during which the cluster of a given node is not linked with any other cluster.
The distributions accumulated in two time intervals, $[0, t_a]$ and $[t_a, t_g(t_b)]$, exhibit power-law decays with exponents $\zeta$ and $\zeta^\prime$, respectively.
The two exponents can be determined in terms of the exponents $\{\tau_s,\sigma_s$\} and $\sigma^\prime(=0)$, which is also true for the values of the remaining critical exponents.

It is noteworthy that the transition point of $m$-BFW model is subject to intrinsic uncertainty thus the conventional finite-size scaling theory in the super-critical region breaks down to some extent. Therefore, we constructed a novel framework for finite-size scaling analysis which adjusts the transition points of each sample for each system size and identify the value characteristic for the system as an average which obeys the scaling laws.

Systems exhibiting HPT constitute a very important class of dynamic percolation problems from epidemic problems to interdependent networks, where the jump in the order parameter represents a dramatic, explosive change of the system. The present study completes the description of the HPT as we have now a theoretical framework for both cluster pruning and cluster merging processes leading to such a transition. In both cases we have two sets of exponents, which are coupled by the scaling relation $\gamma_a+\beta_m=1$ and its counterpart $\gamma_s+\beta_m=1$. We think that the universal mechanism and features we uncovered in this paper can be used for understanding other discontinuous transitions in complex systems such as in synchronization and jamming transitions.

\begin{acknowledgments}
This work was supported by the National Research Foundation of Korea by Grant Nos. NRF-2020R1F1A1061326 (YSC), NRF-2014R1A3A2069005 (BK), KENTECH Research Grant No. KRG2021-01-007, and US Army Research Office Grant No. W911NF-23-1-0087 (RMD), EU Horizon 2020 grant agreement ERC No 810115-DYNASNET.
\end{acknowledgments}

\clearpage
\newpage
\appendix
\onecolumngrid
\renewcommand\thefigure{A\arabic{figure}}
\renewcommand\thetable{A\arabic{table}}
\setcounter{figure}{0}
\setcounter{table}{0}

\section{Critical exponents and scaling relations}  \label{app:exponent}
We present two sets of critical exponents to characterize the critical behaviors of HPT on random complex networks. One set comprising $\{\beta_m, \gamma_m, \nubar_m \}$ is associated with the giant cluster size per node denoted with index $m$. The other set comprising $\{\tau_s, \sigma_s, \beta_s, \gamma_s, \nubar_s\}$ is associated with the size distribution of finite clusters denoted with index $s$. The prime on the exponents represents the exponents in the subcritical regime ($t < t_c$) except for $\zeta^\prime$, specified in the Table below. The definitions of critical exponents are listed in Table.~\ref{tab:exponents}.

\begin{table*}[!hb]
\renewcommand{\arraystretch}{1.2}
\centering
\setlength{\tabcolsep}{1.5em}
\begin{tabular}{clll}
\hline\hline
Exponent & Quantity & Definition & Regime \\
\hline
$\beta_m$ & Giant cluster size per node & $m-m_c \sim (t-t_c)^{\beta_m}$ & $t > t_c$ \\
$\gamma_m^\prime$ & Susceptibility & $\chi_m \sim (t_c-t)^{-\gamma_m^\prime}$ & $t < t_c$ \\
$\gamma_m$ & Susceptibility & $\chi_m \sim (t-t_c)^{-\gamma_m}$ & $t > t_c$ \\
$\nubar_m^\prime$ & Correlation size & $\xi_s\equiv\xi_\ell^d\sim (t_c-t)^{-d\nu_m^\prime}\sim (t_c-t)^{-\nubar_m^\prime}$ & $t < t_c$\\
$\nubar_m$ & Correlation size & $\xi_s\equiv\xi_\ell^d\sim (t-t_c)^{-d\nu_m}\sim (t-t_c)^{-\nubar_m}$ & $t > t_c$\\
\hline
$\tau_s$ & Cluster size distribution at $t_c$ & $n_s(t_c) \sim s^{-\tau_s}$ & At $t_c$\\
$\sigma_s^\prime$ & Characteristic size & $s^* \sim (t_c-t)^{-1/\sigma_s^\prime}$ & $t < t_c$ \\
$\sigma_s$ & characteristic size & $s^* \sim (t-t_c)^{-1/\sigma_s}$ & $t > t_c$\\
$\beta_s$ & Giant cluster size per node & $\beta_s = (\tau_s-2) / \sigma_s$ & $t > t_c$ \\
$\gamma_s^\prime$ & Mean cluster size & $\langle s \rangle \sim (t_c-t)^{-1/\gamma_s^\prime}$ & $t < t_c$ \\
$\gamma_s$ & Mean cluster size & $\langle s \rangle \sim (t-t_c)^{-1/\gamma_s}$ & $t > t_c$ \\
$\nubar_s^\prime$ & Correlation size & $s^* \sim N^{1 / \sigma_s\nubar_s^\prime}$ & $t < t_c$\\
$\nubar_s$ & Correlation size & $s^* \sim N^{1 / \sigma_s\nubar_s}$ & $t > t_c$ \\
\hline
$\zeta$ & Inter-coalescence time distribution & $P_a(z) \sim z^{-\zeta}$ & $t < t_a$ \\
$\zeta'$ & Inter-coalescence time distribution & $P_a(z) \sim z^{-\zeta^\prime}$ & $t_a < t < t_g~ {\rm or}~ t_b$ \\
\hline\hline
\end{tabular}
\caption{
$\chi_m\equiv N(\langle m^2 \rangle - \langle m \rangle^2)$.
The distribution of finite clusters near $t_c$ is expressed as $n_s(t)\sim s^{-\tau_s}f(s/s^*)$ near $t_c$, where $f(x)$ is a scaling function with $f(x)$ being  constant for $x \ll 1$ and $f(x)\sim x^{\tau_s}$ for $x \gg 1$. The characteristic size of finite clusters $s^*\sim (t_c-t)^{-1/\sigma_s^\prime}$. Note that $\sigma_s^\prime$ has different values in $[0, t_a]$ and $[t_a, t_g]$ or $[t_a, t_b]$. $\beta_s$ cannot be directly measured. The mean cluster size of finite clusters $\langle s \rangle \equiv \sum_s^\prime s^2 n_s / \sum_s^\prime s n_s$ where the prime in the summation denotes the sum over finite clusters.
} \label{tab:exponents}
\end{table*}

The values of critical exponents vary depending on the model parameter $g$ ($r$-ER model) or $h$ ($m$-BFW model), representing the suppression strength in the way $1-g$ or $1-h$. The scaling relations among the critical exponents hold independently of the model parameters. The scaling relations are listed in Table.~\ref{tab:relations}.

\begin{table}[!hb]
\renewcommand{\arraystretch}{1.2}
\centering
\setlength{\tabcolsep}{1.5em}
\begin{tabular}{clll}
\hline\hline
Giant cluster & $\nubar_m = 2\beta_m + \gamma_m$ &  & \\
Finite clusters & $\nubar_s = 2\beta_s + \gamma_s$ & $\beta_s = (\tau_s-2) / \sigma_s$ & $\gamma_s = (3-\tau_s)/\sigma_s$\\
Scaling relation between two exponent sets & $\gamma_s + \beta_m = 1$ &  & \\
\makecell{Scaling relation between the exponents of \\ inter-coalescence time and cluster size distributions} & $\zeta^\prime = 4-(\tau_s + \sigma_s^\prime)$ & $\zeta = 4-\tau_s$ & \\
\hline\hline
\end{tabular}
\caption{
Scaling relations among the critical exponents.
} \label{tab:relations}
\end{table}

\section{Difference between the original $r$-ER and the modified $r$-ER models} \label{app:diff_rER}
Here, we explain the difference between the original and modified $r$-ER models in their dynamic rules.
Then, why the modification is necessary to make an HPT occur is discussed.

The original $r$-ER model was proposed to study a discontinuous percolation transition in Ref.~\cite{panagiotou2011explosive}.
In the model, $g$ fraction of nodes that are contained in the smallest clusters are assigned to set $A$ and the remaining fraction of nodes are assigned to set $B$.
As clusters are merged, clusters can be reassigned their sets.
In the original model, the number of nodes assigned to set $A$ is fixed as $\lfloor gN \rfloor$.
As a result, nodes of the cluster on the border are assigned to either set $A$ or $B$.
This assignment rule differs from that of the modified $r$-ER model, where nodes in the cluster on the border of the two sets $A$ and $B$ are regarded as the elements of set $A$.
As a result, when $t$ reaches $t_g$, then $m(t_g)=1-g$, that is, the fractional size of the largest cluster at time $t_g$ becomes the capacity $1-g$ of set $B$, the largest cluster is regarded as the element of set $A$.
Then, the partitions become unified, and the dynamics of cluster merging is governed by the ER dynamics when $t \ge t_c$.

\begin{figure}[!h]
    \centering
    \includegraphics[width=0.60\linewidth]{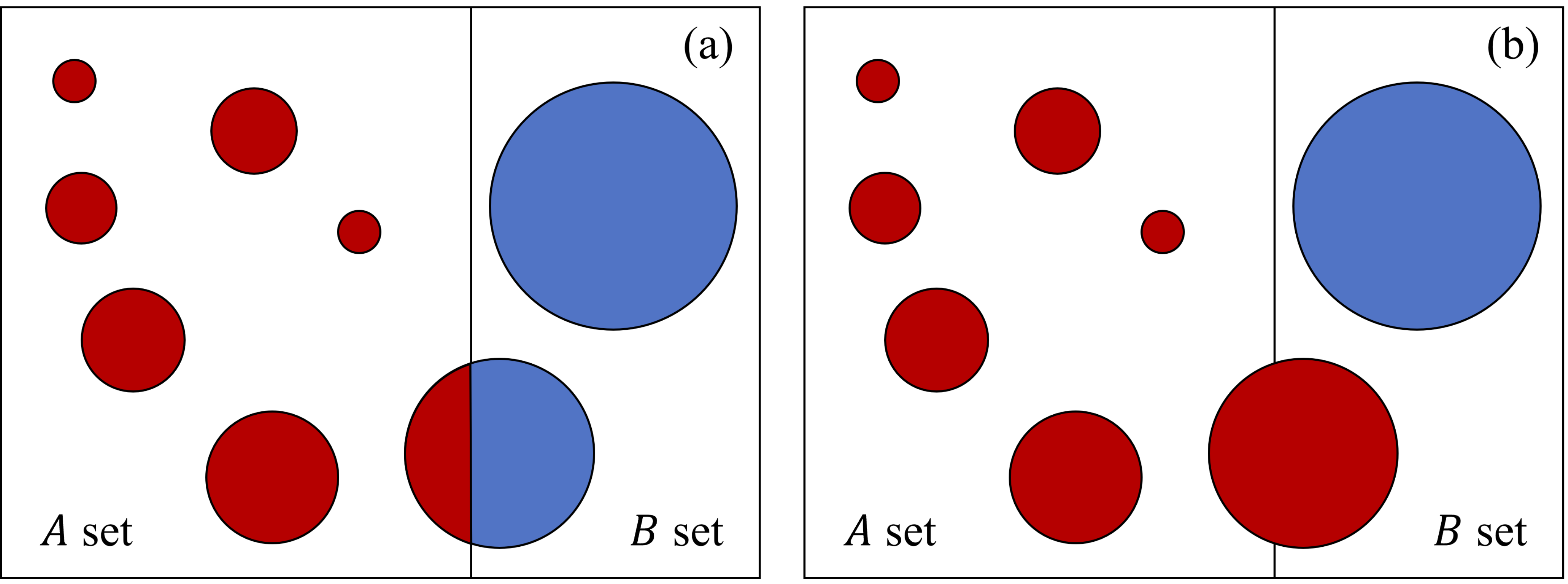}
    \caption{
        Schematic plots of the original $r$-ER model (a) and the modified $r$-ER model (b).
        Circles represent clusters.
        Circles contained in set $A$ ($B$) are marked in red (blue).
    }\label{fig:a1}
\end{figure}

\section{Evolution of the order parameter and the cluster size distribution in the three regimes of the modified $r$-ER model}  \label{app:step}
\begin{figure*}[!htb]
    \centering
    \includegraphics[width=0.99\linewidth]{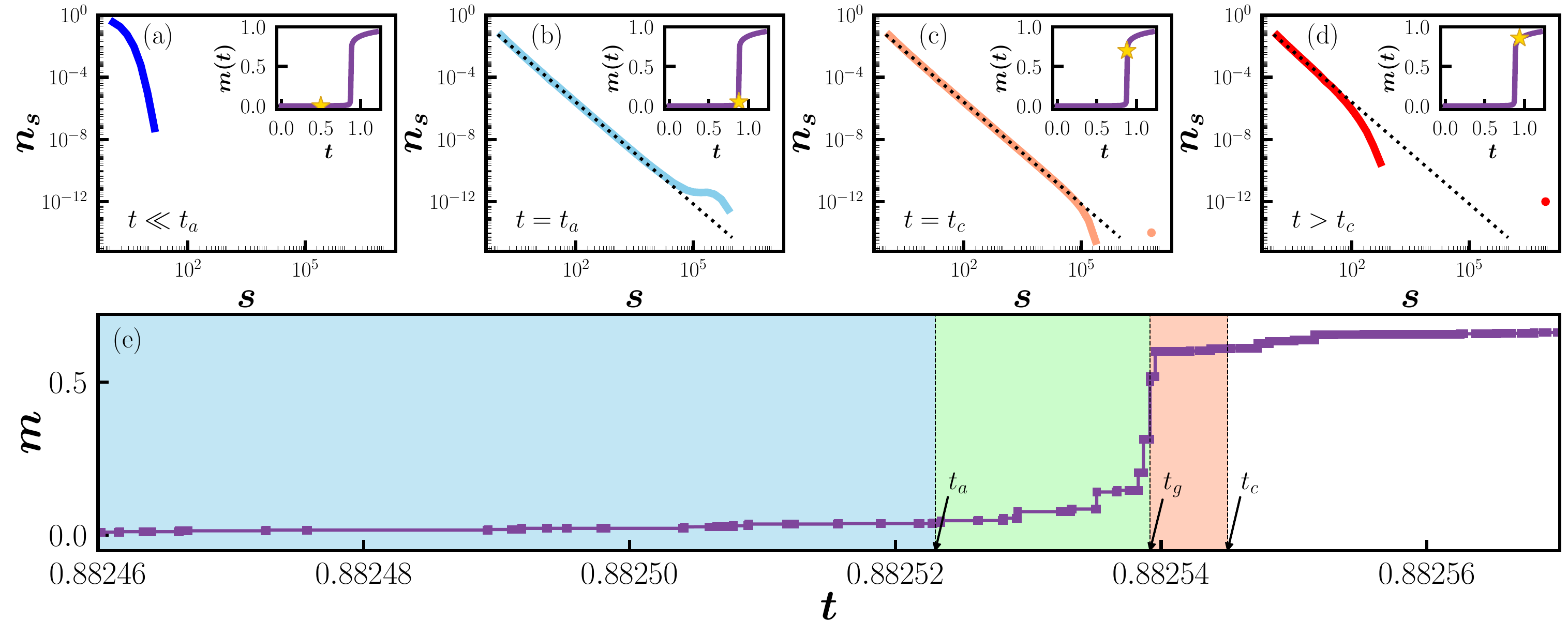}
    \caption{
        Plots of the cluster size distributions of the modified $r$-ER model at (a) $t \ll t_a$, (b) $t=t_a$, (c) $t=t_c$, and (d) $t > t_c$.
        Inset: Schematic plots of the order parameter $m(t)$ vs $t \equiv L/N$ with a point ($\star$) indicating the corresponding time.
        (e) Plot of the order parameter with characteristic points $\{t_a,t_g,t_c\}$ in the evolution of a single sample of the modified $r$-ER model with a system size $N=1.024\times 10^4$ for $g=0.5$.
        Similar plots were presented in Refs.\cite{cho2016hybrid,park2019interevent}
    }  \label{fig:figa3}
\end{figure*}
We plot the evolution of the order parameter and the cluster size distribution of the modified $r$-ER model throughout the three regimes in Fig.~\ref{fig:figa3}.
This plot is shown for the comparison to Fig.~\ref{fig:fig2} of the $m$-BFW model.

\section{Difference between the original and $m$-BFW models}  \label{app:diff_BFW}
\begin{figure}[!htb]
    \centering
    \includegraphics[width=0.60\linewidth]{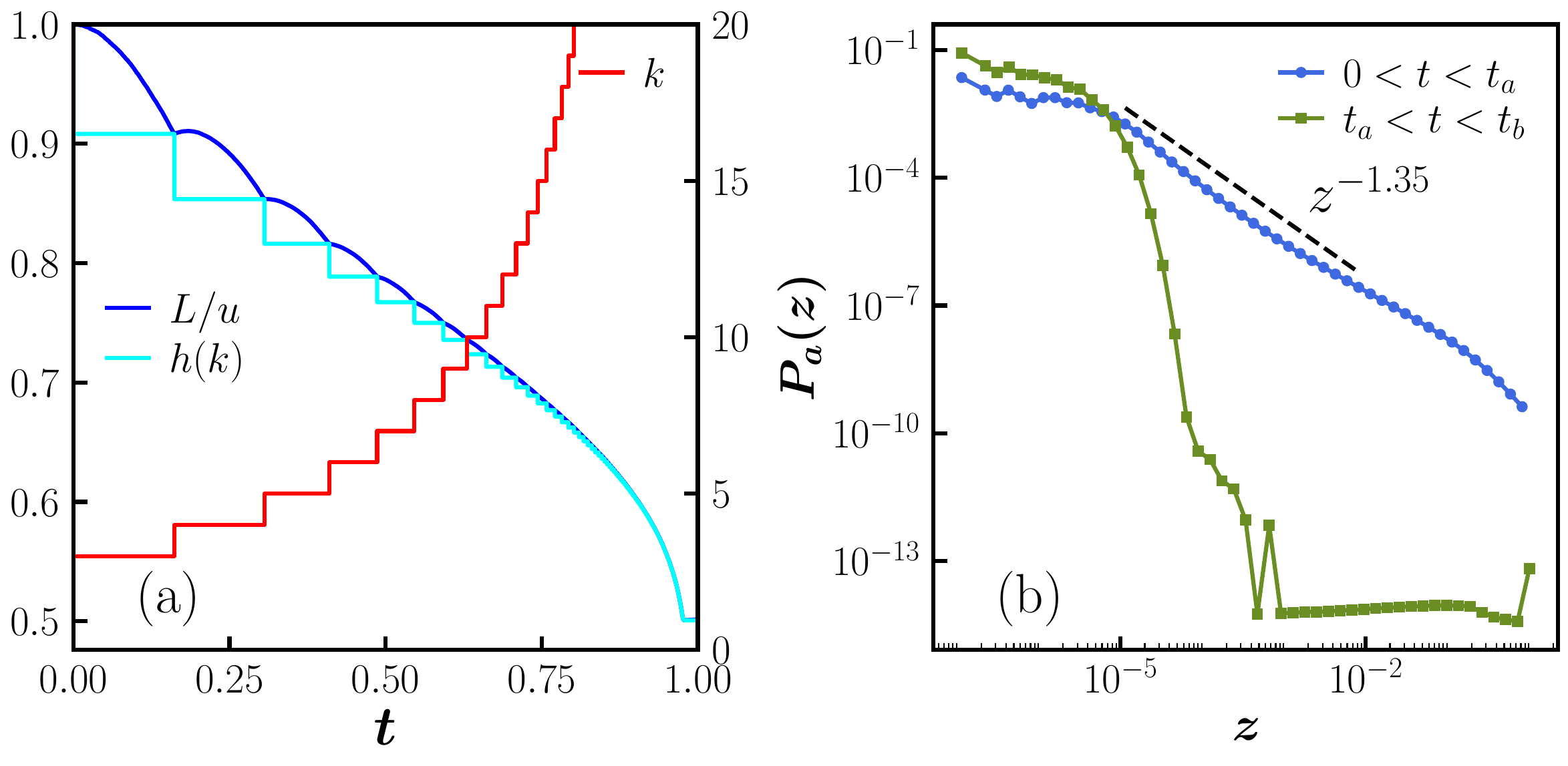}
    \caption{
        (a) Plot of the acceptance rate $L/u$ (left axis) and the limit $k$ of the largest cluster size (right axis) vs time $t$ for the original BFW model.
        $h(k)$ decreases regularly.
        Thus, the limit $k$ increases regularly, and the step widths and heights respectively decrease and increase regularly, respectively.
        The acceptance rate, i.e., $L/u$, also decreases.
        (b) Plots of the age distribution as a function of age $z$.
        The age distribution for $t \in [t_a, t_b]$ does not exhibit power-law behavior, and thus the criticality is not formed for $t > t_c$ and the percolation transition is discontinuous.
    }\label{fig:accept_original}
\end{figure}
The main difference between the original and $m$-BFW models lies in the lower bound $h$ of the acceptance rate $L/u$.
For the $m$-BFW model, $h$ is constant, whereas for the original BFW model, $h(k)$ depends on $k$ as $h(k)=1/2+(2k)^{-1/2}$.
Accordingly, the acceptance rate $L/u$ behaves differently as a function of time $t$ as shown in Fig.~\ref{fig:accept_original}.
We found that $h(k)$ decreases regularly, and thus the upper bound of cluster size denoted as $k$ increases regularly; the step widths and heights respectively decrease and increase regularly.
The acceptance rate, i.e., $L/u$, also decreases.
Accordingly, the age distribution does not show a power-law behavior in the interval $t\in [t_a, t_b]$.

\section{Derivation of $\sigma_s=1$}  \label{app:deriv_sigma}
Let $p_s(\hat t)$ be the probability that a randomly selected node is an element of a cluster of size $s$ at relative time $\hat{t}\equiv t-t_c$.
Then $p_s(0)\sim s^{1-\tau}$ and $\sum_{s=1}^{\prime}p_s(0)=1-m_c$ for HPT.
For $\hat t \geq 0$, the kinetics of the cluster merging of the modified $r$-ER model is governed by the ER dynamics with the initial configuration $p_s(0)$.
For $\hat t > 0$, $p_s(\hat t)$ is analytic with respect to $\hat t$, and thus $p_s(\hat t)$ can be written as
\begin{equation}
    p_s(\hat t)=B_0(s)+B_1(s)\hat t+B_2(s)\hat t^2+\cdots,
\end{equation}
where $B_0(s)=p_s(0)$.
Using the property $\sum_{s=1}^{\prime}B_0(0)=1-m_c$, one can find that $B_1(s)\sim s^{2-\tau}$ and $B_1(s)\sim s^{3-\tau}$, etc.
It is straightforward to derive $B_n(s) \sim s^{n+1-\tau}$ for general $n \geq 0$ using the relation $B_{n+1}(s) \sim sB_{n}(s)$ when $s$ is large.
Therefore, $p_s(\hat{t})$ can be written as $p_s(\hat{t}) = s^{1-\tau}\sum_{n\geq 0} b_n (s\hat{t})^n$ with constant $b_n$.
The derivation of this result was presented in detail in the supplementary information of Ref.~\cite{cho2016hybrid}.
Therefore, $p_s(\hat t)$ can be written in a scaling form, $p_s(\hat t)=s^{1-\tau}f(s\hat t^{1/\sigma_s})$ with $\sigma_s=1$.
This approach can also be used in the $m$-BFW model.

\vfill\eject
\twocolumngrid


%

\end{document}